\documentclass[12pt,letterpaper]{article}
\pdfoutput=1
\usepackage{graphicx}
\usepackage{subfig}
\usepackage{MnSymbol}
\DeclareMathAlphabet\mathbb{U}{msb}{m}{n}

\newcommand{\be}{\begin{equation}}
\newcommand{\ee}{\end{equation}}
\newcommand{\ben}{\begin{displaymath}}
\newcommand{\een}{\end{displaymath}}
\newcommand{\bea}{\begin{eqnarray}}
\newcommand{\eea}{\end{eqnarray}}
\newcommand{\bean}{\begin{eqnarray*}}
\newcommand{\eean}{\end{eqnarray*}}

\newcommand{\bX}{\ensuremath{\bar{X}}}

\newcommand{\ads}[1]{\mbox{${AdS}_{#1}$}}

\newcommand{\eg}{{\it e.g.}}
\newcommand{\ie}{{\it i.e.}}

\newcommand{\tr}{\mbox{Tr}}

\newcommand{\commentout}[1]{}

\newcommand{\beq}{\begin{equation}}
\newcommand{\eeq}{\end{equation}}
\newcommand{\beqr}{\begin{displaymath}}
\newcommand{\eeqr}{\end{displaymath}}
\newcommand{\beqa}{\begin{eqnarray}}
\newcommand{\eeqa}{\end{eqnarray}}
\newcommand{\beqar}{\begin{eqnarray*}}
\newcommand{\eeqar}{\end{eqnarray*}}
\newcommand{\cN}{{\cal N}}

\newcommand{\cO}{{\cal O}}

\newcommand{\half}{\ensuremath{\frac{1}{2}}}

\newcommand{\bz}{\ensuremath{\bar{z}}}

\newcommand{\N}[1]{\ensuremath{\cN=#1}}

\newcommand{\cotan}{\ensuremath{\mbox{cotan}}}
\newcommand{\cU}{\ensuremath{\mathcal{U}}}

\renewcommand{\Re}{\ensuremath{\mathrm{Re}}}
\renewcommand{\Im}{\ensuremath{\mathrm{Im}}}

\begin{document}

\title{\LARGE \bf Wilson loops and minimal area surfaces in hyperbolic space}

\author{Martin Kruczenski\thanks{E-mail: \texttt{ markru@purdue.edu}}\\
        Department of Physics and Astronomy, Purdue University,  \\
        525 Northwestern Avenue, W. Lafayette, IN 47907-2036. }

\maketitle

\begin{abstract}

The AdS/CFT correspondence relates Wilson loops in \N{4} SYM theory to minimal area surfaces in AdS space. If the loop is a plane curve the minimal 
surface lives in hyperbolic space $\mathbb{H}_3$ (or equivalently Euclidean \ads{3} space). We argue that finding the area of such extremal surface can be easily done if we solve the following problem: 
given two real periodic functions $V_{0,1}(s)$, $V_{0,1}(s+2\pi)=V_{0,1}(s)$ a third periodic function $V_2(s)$ is to be found such that all solutions to the equation 
$-\partial_s^2 \phi + \left[V_0+\half(\lambda+\frac{1}{\lambda}) V_1 + \frac{i}{2} (\lambda-\frac{1}{\lambda}) V_2\right] \phi=0$ are anti-periodic in $s\in[0,2\pi]$ for any value of $\lambda$. 
This problem is equivalent to the statement that the monodromy matrix is trivial. It can be restated as that of finding a one complex parameter family of curves $X(\lambda,s)$ where $X(\lambda=1,s)$ 
is the given shape of the Wilson loop and such that the Schwarzian derivative $\{X(\lambda,s),s\}$ is meromorphic in $\lambda$ with only two simple poles. We present a formula for the area in terms of the functions $V_{0,1,2}$ and discuss solutions to these equivalent problems in terms of theta functions. Finally, we 
also consider the near circular Wilson loop clarifying its integrability properties and rederiving its area using the methods described in this paper. 

\end{abstract}

\clearpage
\newpage



\section{Introduction}
\label{intro}

 The most fundamental operator in a gauge theory is the Wilson loop. It can distinguish a confining from a non-confining phase, it determines the quark/anti-quark potential and by expanding it at small distances one can obtain the expectation value of any local gauge invariant operator.  Thus, one of the first and most important results of the AdS/CFT correspondence \cite{malda} was to give an alternative computation of the Wilson loops at strong coupling in \N{4} SYM by relating it to a minimal area surface in \ads{} space\cite{MRY}. Much work has been devoted to the computation of explicit examples of Wilson loops. For Euclidean curves, the most studied case is the circular Wilson loop \cite{cWL} although another cases have been considered \cite{WLref}. In the case
 of Minkowski signature the light-like cusp \cite{cusp} turns out to be particularly interesting because of its relation to scattering amplitudes \cite{scatampl}. To find solutions in all those cases it is important to exploit the integrability properties of the
 equations of motion which are the same as those of the closed string. Recently, in the case of closed Euclidean plane Wilson loops (with constant scalar) an infinite parameter family of analytical solutions was found using Riemann theta functions \cite{IKZ,KZ} following results form the mathematical literature \cite{BB, BBook} and from previous results for closed strings 
 \cite{ClosedStrings}. This integrability construction for the Wilson loop was further discussed in \cite{Janik} and also in 
 \cite{WLint}. More recently certain integrability
 properties of the near circular Wilson loop were explained in \cite{Cagnazzo}. 

 In this paper we study more in detail the integrable structure that allows the computation of those surfaces. Integrability of the string sigma model implies the existence of an infinite number of conserved quantities given by the holonomy of a certain flat current along a non-trivial loop. A standard application of integrability is to use the conserved quantities to determine the evolution of a string once a complete set of initial data is given, namely the initial position and velocity of the string. Instead, in the Euclidean case considered in this paper, the world-sheet has the topology of a disk and all loops are trivial implying that all the conserved quantities vanish. Equivalently, instead of a complete set of boundary data we are only given half of it, in this case the position. If we choose the other half, namely the radial derivative, arbitrarily, the solution we find will not correspond to a surface that closes smoothly. The condition of vanishing charges is precisely  equivalent to the condition that the surface closes smoothly and allows to determine the other half of the boundary data in order to set up the computation as an initial problem.  
 Therefore, we argue that the vanishing of the holonomy is the defining property of the Wilson loop and should be used as the basis of constructing the surface and computing the area. The problem is closely analogous to the one of solving the Laplace equation $\partial\bar{\partial} \phi(z,\bar{z})=0$ on a disk $|z|\le1$ given the value at the boundary $|z|=1$. If we know the function and its radial derivative at the boundary then, using the Laplace equation, all higher radial derivatives are determined and the solution can be reconstructed, namely we would have an initial value problem. But we are only given the value of the function. If we choose arbitrarily the normal derivative, continuing the function to the interior will lead to a singularity. The condition for the solution to be smooth is that the normal derivative and the function are related by a dispersion relation which expresses the vanishing of all conserved quantities, in this case $q_{n} = \oint_{|z|=1} dz z^n \partial \phi=0$, $\forall n\in\mathbb{Z}_{\ge 0}$. Equivalently, $q_{n\ge0}=0$ establishes that $\partial \phi$ is holomorphic in the disk. Moreover, the problem of solving the Laplace equation is directly related to the problem of finding a minimal area surface ending on a given contour in flat space. That problem is obviously related to the one we discuss in this paper and for that reason we summarize it briefly in the Appendix. 

 This paper is organized as follows. In the next section we introduce the notation and define the problem. In the following section we show that given two real function at the boundary of the disk, the area can be easily computed. In analogy with the Laplace equation, one of those functions is given by the data of the problem whereas the other follows from a consistency condition. This is analyzed in the subsequent section where the consistency condition is seen to be that all conserved quantities vanish. 
This problem is equivalent to the following one: given two real periodic functions $V_{0,1}(s)$, $V_{0,1}(s+2\pi)=V_{0,1}(s)$ a third periodic function $V_2(s)$ is to be found such that all solutions to the equation 
$-\partial_s^2 \phi + \left[V_0+\half(\lambda+\frac{1}{\lambda}) V_1 + \frac{i}{2} (\lambda-\frac{1}{\lambda}) V_2\right] \phi=0$ are anti-periodic in $s\in[0,2\pi]$ for any value of $\lambda$. 
Equivalently, one can try to find a one complex parameter family of curves $X(\lambda,s)$ such that $X(\lambda=1,s)$ is the shape of the Wilson loop
 and the Schwarzian derivative $\{X(\lambda,s),s\}$ is meromorphic in $\lambda$ with only two simple poles. Finding the relation between those problems and the minimal area problem is the main result of this paper. Unfortunately we were not able
to find a straight-forward and general analytic or numerical solution to those problems leaving that for future work. Instead we find particular solutions based on theta-functions and also perturbatively around the circular solution. Those cases
reproduced known solutions and provide an illustration of the techniques described in this paper..

\section{Statement of the problem and notation}

 Consider Euclidean \ads{3} or equivalently hyperbolic $\mathbb{H}_3$ space parameterized by a real $Z$ and a complex $X$ coordinate with metric given by
\beq
 ds^2 = \frac{dZ^2 + dX d\bar{X}}{Z^2}\ ,
 \label{a1}
\eeq
and an $\mathbb{R}^2\equiv\mathbb{C}$ conformal boundary parameterized by $X$ and located at $Z=0$. 
We are looking for a minimal area surface in this space ending on a given boundary curve $X(s)$. More precisely, given a complex coordinate
\beq
z=\sigma+i\tau = r\, e^{i\theta}\ ,
 \label{a2}
\eeq
we look for a minimal area embedding $X(r,\theta)$, $Z(r,\theta)$ of the unit disk $|z|=r\le 1$ into $\mathbb{H}_3$ such that $Z(r=1,\theta)=0$ and $X(r=1,\theta) = X(s(\theta))$ for the given curve $X(s)$. 
At this point we allow for a boundary reparameterization  $s(\theta)$ since we want to preserve the freedom to choose conformal coordinates in the unit disk. 

\begin{figure}
\centering
\includegraphics[width=10cm]{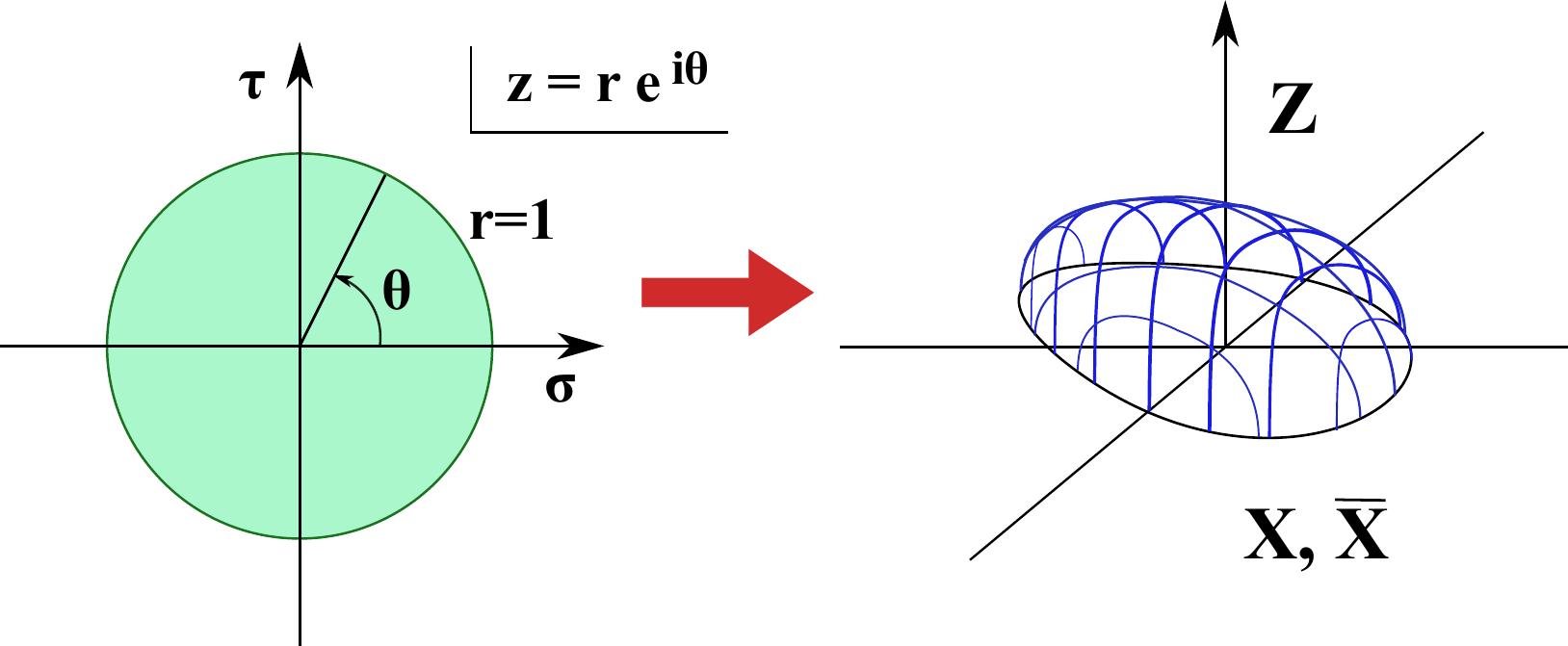}
\caption{The unit disk $|z|<1$ on the left is mapped to a surface $X(z,\bar{z})$, $\bar{X}(z,\bar{z})$, $Z(z,\bar{z})$ on the right. The objective is to find the surface of minimal area ending on a given boundary contour $X(s)$, namely $Z(r=1,\theta)=0$ and $X(r=1,\theta)=X(s(\theta))$ for some reparameterization $s(\theta)$. }
\label{surface}
\end{figure}

 To write the condition of minimal area, it is convenient to describe $\mathbb{H}_3$ as a subspace of $\mathbb{R}^{3,1}$ defined by the constraint
\beq
  X_0^2 - X_1^2 - X_2^2 - X_3^2 = 1\ ,
\label{Xsq1}
\eeq
with an obvious  $SO(3,1)\equiv SL(2,\mathbb{C})$ global invariance. This space has an $S^2$ boundary at infinity. The relation to the Poincare coordinates is straight-forward:
\beq
 Z = \frac{1}{X_0-X_3}, \ \ \ X = \frac{X_1+iX_2}{X_0-X_3}, \ \ \ \bar{X} = \frac{X_1-iX_2}{X_0-X_3}\ .
  \label{a3}
\eeq
 The area in the conformal parameterization of the surface is given by
\beq
 S = \half \int \left( \partial X_\mu \bar{\partial} X^\mu + \Lambda (X_\mu X^\mu -1)\right) \ d\sigma\, d\tau
   \label{action}
\eeq
where $\Lambda$ is a Lagrange multiplier, the $\mu$ indices are raised and lowered with the $\mathbb{R}^{3,1}$ metric and $\partial$, $\bar{\partial}$ denote derivative with respect to $z$, $\bar{z}$.
A minimal area surface is given by real functions $X_\mu(z,\bar{z})$ obeying the equations:
\beq
  \partial \bar{\partial} X_\mu = \Lambda X_\mu\ ,
\label{eomX}
\eeq
where $\Lambda$, is given by 
\beq
\Lambda = -\partial X_\mu \bar{\partial} X^{\mu}. \label{Lcomp}
\eeq
Finally, we should additionally impose the Virasoro or conformal constraints which read
\beq
 \partial X_\mu \partial X^\mu = 0 = \bar{\partial} X_\mu \bar{\partial} X^\mu.
\label{Vc1}
\eeq
These equations can be rewritten using the matrix
\beq
\mathbb{X} = \left(\begin{array}{cc} X_0+X_3 & X_1-i X_2 \\ X_1 + i X_2 & X_0-X_3 \end{array} \right) = X_0 + X_i \sigma^i \ ,
 \label{a4}
\eeq
where $\sigma^i$ denote the Pauli matrices. Notice also that the Poincare coordinates are simply given by
\beq
Z = \frac{1}{\mathbb{X}_{22}} , \ \ \ X =\frac{\mathbb{X}_{21}}{\mathbb{X}_{22}}.
\label{PoinX}
\eeq
The matrix $\mathbb{X}$ satisfies
\beq
\mathbb{X}^\dagger = \mathbb{X} , \ \ \det \mathbb{X} = 1, \ \ \ \partial\bar{\partial} \mathbb{X}=\Lambda \mathbb{X}, \ \ 
\det(\partial\mathbb{X})=0=\det(\bar{\partial}\mathbb{X})\ ,
 \label{a5}
\eeq
as follows from the definition of $\mathbb{X}$, the constraint (\ref{Xsq1}), the equations of motion (\ref{eomX}) and the Virasoro constraints (\ref{Vc1}). 
We can solve the constraint $\mathbb{X}^\dagger=\mathbb{X}$ by writing 
\beq
\mathbb{X}=\mathbb{A}\mathbb{A}^\dagger, \ \ \ \ \det \mathbb{A}=1 , \ \ \ \ \mathbb{A} \in SL(2,\mathbb{C}).
 \label{a6}
\eeq
The equations of motion have a global $SL(2,\mathbb{C})\equiv SO(3,1)$ symmetry under which
\beq
\mathbb{X} \rightarrow U\mathbb{X}U^\dagger, \ \ \ \mathbb{A}\rightarrow U\mathbb{A}, \ \ \ U\in SL(2,\mathbb{C}).
 \label{a7}
\eeq
In the new variable there is an $SU(2)$ gauge symmetry
\beq
 \mathbb{A} \rightarrow \mathbb{A} \cU, \ \ \ \cU(z,\bar{z})\in SU(2)\ ,
  \label{a8}
\eeq
since this leaves $\mathbb{X}$ invariant. It is useful to define the current
\beq
 j = \mathbb{A}^{-1} d\mathbb A = J dz + \bar{J} d\bar{z} \ ,
\label{Jdef}
\eeq
which is invariant under the global symmetry and, under the local symmetry transforms as
\beq
j \rightarrow \cU^{\dagger} j \,\cU + \cU^\dagger d \cU\, .
 \label{a9}
\eeq
 Using the local symmetry and the equations of motion, this current can be put in the form (see for example \cite{IKZ,KZ})
\beq
J=\left(\begin{array}{ccc}-\half\partial\alpha &\ & f(z) e^{-\alpha}\\ && \\ \lambda e^{\alpha} &&  \half \partial\alpha  \end{array}\right), \ \ \ \ 
\bar{J}=\left(\begin{array}{ccc}\half \bar{\partial}\alpha &\ & \frac{1}{\lambda} e^{\alpha}\\ && \\-\bar{f}(\bar{z})e^{-\alpha} && -\half \bar{\partial}\alpha  \end{array}\right) .
 \label{a10}
\eeq
where $f(z)$ is an arbitrary holomorphic function $\bar{\partial}f=0$, $\alpha$ is a real function in the unit disk $|z|=r<1$ such that  
\beq
  \partial\bar{\partial}\alpha = e^{2\alpha} +f \bar{f} e^{-2\alpha} , \label{alphaeqn} 
\eeq
and $\lambda$ in eq.(\ref{a10}) is an arbitrary parameter known as the spectral parameter. 
 Under these conditions, the current 
\beq
 j =J dz + \bar{J} d\bar{z} \ ,
  \label{a11}
\eeq
satisfies
\beq
 dj + j\wedge j =0\ ,
  \label{a12}
\eeq
for all values of $\lambda$. As an aside, notice also the validity of the reality condition
\beq
 \left[j\left(-\frac{1}{\bar{\lambda}}\right)\right]^\dagger = - j(\lambda)\ .
\eeq
Therefore, a way to solve the equations of motion is to first solve eq.(\ref{alphaeqn}) then plug $\alpha$ into the definitions for $J$, $\bar{J}$, namely eq.(\ref{a10}), and solve for $\mathbb{A}$:
\beqa
  \partial \mathbb{A} &=& \mathbb{A} J   \ ,    \label{A1eqn} \\
  \bar{\partial} \mathbb{A} &=& \mathbb{A} \bar{J} .  \label{A2eqn}
 \eeqa
  Finally, the surface is determined as $\mathbb{X}=\mathbb{A}\mathbb{A}^\dagger$. This procedure, in fact, defines a one parameter family of surfaces, one for each value of $\lambda$. 
  The only ones that are solutions of the equations of motion are those corresponding to $|\lambda|=1$ and they turn out to have all the same area. For concreteness, we take the solution
  we are interested in to be the one for $\lambda=1$.   
  
  In any case, the equation for $\alpha$ is non-linear but the ones for $\mathbb{A}$  are linear since $J$, $\bar{J}$ are known once $\alpha$ is known. This is the main idea of the Pohlmeyer reduction \cite{Pohlmeyer} which we rederive here as it applies to our  particular problem. Similar considerations in the context of string theory are well-known, for example see \cite{ClosedStrings,WLint} and \cite{Hoare:2012nx}.
  
 Notice that, $\mbox{Tr}J=\mbox{Tr}\bar{J}=0$ implies that $\det \mathbb{A}$ is constant independent of  $z,\bar{z}$. Since we need $\det \mathbb{A}=1$ we can just normalize 
 $\mathbb{A}$ dividing by a constant. Furthermore it is convenient to write
 \beq
 \mathbb{A} = \left(\begin{array}{cc}\psi_1 & \psi_2\\ \tilde{\psi}_1 & \tilde{\psi}_2\end{array}\right)\ ,
  \label{a13}
 \eeq
where the vectors $\psi=(\psi_1,\psi_2)$ and $\tilde{\psi}=( \tilde{\psi}_1, \tilde{\psi}_2)$  are linearly independent and satisfy
\beq
\partial \psi = \psi J, \ \ \ \bar{\partial} \psi = \psi \bar{J}\ , \label{p12eq}
\eeq
and the same for $\tilde{\psi}$. They have to be linearly independent and are normalized such that the (constant) determinant $\det\mathbb{A}=\psi_1 \tilde{\psi}_2 -\psi_2 \tilde{\psi}_1=1$. There is still a certain ambiguity in choosing $\psi$, $\tilde{\psi}$ but those correspond to
$SL(2,\mathbb{C})\equiv SO(3,1)$ transformations of $\mathbb{X}$. 

 In \cite{IKZ, KZ} it was shown, following \cite{BB,BBook} how to find an infinite parameter family of solutions to the equations in terms of theta functions. 
In what follows we are going to discuss how the integrability properties of the equations of motion can be exploited to further understand this problem and use those
solutions as example.  

\section{Solution given the parameterization $X(\theta)$}

 A mentioned before, the boundary curve as data is given as a function $X(s)$, $s\in[0,2\pi]$ and is related by an unknown reparameterization $s(\theta)$ to the boundary value $X(r=1,\theta)$ of the function $X(z,\bar{z})$ 
in the conformal parameterization of the disk. In this section we are going to assume that such reparameterization is known and show how the data $X(\theta)$ allows for a simple computation of the area. At the end of the 
section we rewrite the formulas in terms of the parameterization $X(s)$ at the expense of introducing an unknown function $V_2(s)$. In the next section we discuss how to determine such function. 
We start by studying the properties of the function $\alpha$ and how it can be reconstructed from certain boundary data. Later we show that the boundary data can be obtained from $X(\theta)$ and that, from there, the area simply follows.  

\subsection{Expansion near the boundary of the disk}

The function $\alpha$ determines the metric induced on the surface as
\beq
 ds^2 = 4 e^{2\alpha}dz d\bar{z}\ .
  \label{b1}
\eeq
 Since the induced metric diverges at the boundary $Z=0$ of $\mathbb{H}_3$ (due to the factor $\frac{1}{Z^2}$ in the metric), it follows that 
\beq
 \alpha(r,\theta) \rightarrow \infty, \ \ \ \mbox{when} \ \ \ \ r\rightarrow 1\ .
  \label{b2}
\eeq
 Consider now the equation (\ref{alphaeqn}) for $\alpha$
\beq
 \partial\bar{\partial}\alpha = e^{2\alpha} +f \bar{f} e^{-2\alpha} \ ,
  \label{b3}
\eeq
for a given analytic function $f(z)$ in the unit disk. To expand the solution near the boundary it is further convenient to define the coordinate
 \beq
 \xi = 1-r^2\ ,
  \label{b4}
 \eeq
that vanishes at $r=1$. Expanding near $\xi=0$ we find the solution
\beq
 \alpha = -\ln \xi + \beta_2(\theta) (1+\xi)\xi^2 + \beta_4(\theta) \xi^4 + \cO(\xi^5) \ ,
  \label{b5}
\eeq
 The arbitrary function $\beta_2(\theta)$ completely determines the solution since all the higher coefficients are algebraic functions of $f(z)$, $\beta_2$ and its derivatives. For example
 \beq
 \beta_4(\theta) = \frac{1}{10} |f(e^{i\theta})|^2 + \frac{1}{5} \beta_2^2 - \frac{1}{40} \beta''_2 + \frac{9}{10} \beta_2\ ,
  \label{b6}
\eeq
 The function $\beta_2(\theta)$ can also be defined as
 \beq
  \beta_2(\theta) = \left. \frac{1}{6} e^{2i\theta} (\partial^2 \alpha-(\partial\alpha)^2)\right|_{r\rightarrow 1}\ .
   \label{b7}
 \eeq
Since the function $f(z)$ is completely determined by its boundary value $f(r=1,\theta)$, the functions $\alpha(r,\theta)$ and $f(z)$ are completely
determined by two functions $f(r=1,\theta)$ and $\beta_2(\theta)$. This data however is redundant, if we are given $f(r=1,\theta)$ choosing $\beta_2(\theta)$ arbitrarily should lead to a singular 
solution in similar fashion as it happens for the Laplace equation if we give the value of the function and the normal derivative. Let us assume for the moment that we know those functions 
and want to find the shape of the curve where the corresponding surface ends.

\subsection{Shape of the Wilson loop}

Having computed the expansion for $\alpha$, the expansion for $J$ and $\bar{J}$ immediately follows. Having $J$, $\bar{J}$ we have to solve the linear problem
\beq
  d (\psi_1,\psi_2) = (\psi_1,\psi_2) j\ .
   \label{b8}
\eeq
 Given two linearly independent solutions $(\psi_1,\psi_2)$ and $(\tilde{\psi}_1,\tilde{\psi}_2)$  we construct
\beq
  \mathbb{A} = \left(\begin{array}{cc}\psi_1 & \psi_2 \\ \tilde{\psi}_1 & \tilde{\psi}_2 \end{array} \right) \ ,
   \label{b9}
\eeq
 and then 
 \beq
 \mathbb{X} = \mathbb{A} \mathbb{A}^\dagger\ .
  \label{b10}
 \eeq
The shape of the boundary is given by\footnote{Some formulas are more conveniently written in terms of $\bar{X}$ instead of $X$ but it is completely equivalent since they are just conjugate of each other.}
\beq
\bX = \frac{\mathbb{X}_{12}}{\mathbb{X}_{22}} = \frac{\psi_1\tilde{\psi}_1^*+\psi_2\tilde{\psi}_2^*}{\tilde{\psi}_1\tilde{\psi}_1^*+\tilde{\psi}_2\tilde{\psi}_2^*} 
= \frac{\psi_1}{\tilde{\psi}_1}\left(\frac{1+\frac{\psi_2\tilde{\psi}_2^*}{\psi_1\tilde{\psi_1}^*}}{1+\frac{\tilde{\psi}_2\tilde{\psi}_2^*}{\tilde{\psi}_1\tilde{\psi}_1^*}}\right)\ ,
 \label{b11}
\eeq 
 and the functions should be evaluated at the boundary of the disk.  
 Using the expansion obtained for $\alpha$ it follows that, at leading order
\beqa
  \partial_\xi \psi_1 &\simeq& -\frac{\lambda}{2\xi} e^{i\theta} \psi_2 \ ,    \label{b12} \\
  \partial_\xi \psi_2 &\simeq& -\frac{1}{2\lambda} e^{-i\theta} \psi_1   \ .  \label{b13}
\eeqa
 Defining
\beq
  H = \frac{\psi_1}{\psi_2}\ ,
   \label{b14}
\eeq
it follows that
\beq
 \partial_\xi H \simeq -\frac{\lambda}{2\xi} e^{i\theta} + \frac{1}{2\lambda\xi} e^{-i\theta} H^2\ ,
 \label{b15}
\eeq
 the only possible solution is that 
\beq
 H(\xi\rightarrow 0) = \pm \lambda e^{i\theta}\ .
 \label{b16}
\eeq
 Replacing in the value for $X$, namely eq.(\ref{b11}) we find the contour of the Wilson loop as
\beq
    \bX = \frac{\psi_1}{\tilde{\psi}_1} , \ \ \  X = \frac{\psi^*_1}{\tilde{\psi}^*_1} \ .
  \label{b17}
\eeq
Since $\psi_{1}$, $\tilde{\psi}_1$ solve a linear problem with a current holomorphic in $\lambda$, we find the very important property that the boundary contour $\bX(\theta,\lambda)$ is a holomorphic function of $\lambda$ (and $X(\theta,\bar{\lambda})$ is anti-holomorphic). More precisely, $\bX(\theta,\lambda=1)$ gives the shape of the Wilson loop and the solution of the linear problem extends that to a holomorphic, one parameter family of contours $\bX(\theta,\lambda)$. As mentioned before, when $|\lambda|=1$, those contours define minimal area surfaces with the same area as the original one. Finally, notice also that we can take two different solutions of the linear problem and get the contour
\beq
 \tilde{X} = \frac{A\psi^*_1+B\tilde{\psi}^*_1}{C\psi^*_1+D\tilde{\psi}^*_1} = \frac{AX +B}{CX+D}, \ \ \ \ A,B,C,D\in\mathbb{C}, \ \ \ \ AB-CD=1\ ,
  \label{b18}
\eeq
 namely a global conformal transformation of the first one. Since the theory is conformal both are equivalent. Therefore, if we know the solutions to the linear problem near the boundary we can reconstruct the shape of the Wilson loop. As we discuss next, to find it, it is not necessary to solve the linear problem inside the disk, we only need to solve a differential equation along the boundary. So, given the boundary curve $X(\theta)$ in the correct parameterization, we can determine  $f(r=1,\theta)$ and $\beta_2(\theta)$ thus completely determining $f(z)$ and $\alpha(r,\theta)$. 

 \subsection{Solution of the linear problem along the boundary}
 
 At fixed radius on the disk we can solve the linear problem
\beq
 (\partial_{\theta} \psi_1, \partial_\theta \psi_2) = (\psi_1, \psi_2)\, J^\theta \ , 
\label{b18b}
\eeq
with
\beq
 J^\theta= \left(\begin{array}{cc} J^\theta_{11} & J^\theta_{12} \\ J^\theta_{21} & J^\theta_{22} \end{array} \right)
 =i \left(\begin{array}{cc} -\half r\partial_r\alpha & zfe^{-\alpha}-\frac{1}{\lambda}\bar{z}e^\alpha \\ z\lambda e^{\alpha}+\bar{z}\bar{f}e^{-\alpha}  & \half r\partial_r\alpha \end{array} \right)\ .
  \label{b19}
\eeq
 Simple algebra leads to an equation just for $\psi_1$: 
\beq
 \partial_\theta^2 \psi_1 -\partial_\theta \psi_1 (\tr J^{\theta}+\partial_\theta \ln J^\theta_{21}) + \psi_1 (\det J^\theta + J^{\theta}_{11}\partial_\theta\ln\frac{J^{\theta}_{21}}{J^{\theta}_{11}}) =0 \ .
  \label{b20}
\eeq 
 Taking into account that $\tr J=0$ and defining 
 \beq
  \chi(\theta) = \frac{1}{\sqrt{J^\theta_{21}}} \psi_1 \ ,
   \label{b21}
 \eeq
we get
\beq
 -\partial^2_\theta \chi +V(\theta,r) \chi =0\ ,
  \label{b22}
\eeq
with
\beq
 V(\theta,r) = -\half \partial_\theta^2 \ln J^{\theta}_{21} + \frac{1}{4}(\partial_\theta\ln J^{\theta}_{21})^2 - \det J^{\theta} - J^{\theta}_{11} \partial_\theta \ln \frac{J^{\theta}_{21}}{J^{\theta}_{11}} \ .
  \label{b23}
\eeq
Although this is valid for any value of $r$ we want to study what happens as $r\rightarrow 1$. In that limit the potential is finite and equal to
\beq
 V(\theta,r=1) = V(\theta) = -\frac{1}{4} + 6\beta_2(\theta) - f(\theta) \lambda e^{2i\theta} + \frac{1}{\lambda} e^{-2i\theta} \bar{f}(\theta)\ .
  \label{b24}
\eeq
Also, near the boundary, 
\beq
 J^\theta_{21} = i \frac{\sqrt{1-\xi}}{\xi} \lambda \, e^{i\theta} + \cO(\xi) \ , 
\eeq
 and therefore in eq.(\ref{b21}),  the factor $\frac{1}{\sqrt{J_{21}}}$  makes $\chi(\theta)$ antiperiodic instead of periodic as $\psi_1(\theta)$. 
 Given two linearly independent solutions of this equation $\chi_1$ and $\tilde{\chi}_1$ the shape of the Wilson loop is given by
\beq
 \bX = \frac{\chi_1}{\tilde{\chi}_1} \ .
  \label{b25}
\eeq
 Now we can use a well-known property of the Schwarzian derivative (that follows by simple computation) to obtain
\beq
 \{ \bX,\theta\} = \{\frac{\chi_1}{\tilde{\chi}_1},\theta\} = -2 V(\theta) \ .
  \label{b26}
 \eeq
Namely, the Schwarzian derivative of the shape of the contour is given by
\beq
 \{ \bX^\lambda,\theta\} = \half -12 \beta_2(\theta) + 2 f(\theta) \lambda e^{2i\theta} - \frac{2}{\lambda} e^{-2i\theta} \bar{f}(\theta)\ .
  \label{b27}
\eeq
 If we take $\lambda=1$ as defining the Wilson loop of interest then we have the very simple relation
\beqa
\mathrm{Re} \{ \bX,\theta\} &=& \half -12 \beta_2(\theta) \ ,      \label{b28}\\
 \mathrm{Im} \{ \bX,\theta\} &=&  4\mathrm{Im} \left[ f(\theta) e^{2i\theta} \right]  \ . \label{b29}
\eeqa
 Summarizing, given the boundary contour $X(\theta)$ we can compute the Schwarzian derivative and from there we get the necessary boundary data
\beqa
  \beta_2(\theta) &=& \frac{1}{24} - \frac{1}{12} \mathrm{Re} \{ \bX,\theta\}    \ ,         \label{b30}\\
  \mathrm{Im} \left[ f(\theta) e^{2i\theta} \right] &=& \frac{1}{4}  \mathrm{Im} \{ \bX,\theta\} \ . \label{b31}
\eeqa
Since $z^2 f(z)$ is holomorphic we can reconstruct $z^2 f(z)$ in the interior of the disk from the imaginary part at the boundary using the formula 
\beq
 \left. \mathrm{Re} (z^2 f(z)) \right|_{z=e^{i\theta_0}} = \frac{1}{2\pi} \strokedint \mathrm{Im}(z^2 f(z))\, \cotan\left(\frac{\theta-\theta_0}{2}\right)\ .
  \label{b32}
\eeq
This means that, if we are given $\{\bX,\theta\}$ for $\lambda=1$ we can reconstruct it for any value of $\lambda$ as
\beqa
 \{ \bX^\lambda,\theta \} &=& \{\bX,\theta \} + \frac{i}{2}\left(\lambda+\frac{1}{\lambda}-2 \right) \mathrm{Im} \{\bX,\theta\}     \label{b33}\\
 &&+\frac{1}{4\pi} \left(\lambda-\frac{1}{\lambda}\right) \strokedint \mathrm{Im} (\{\bX,\theta'\})\, \cotan\left(\frac{\theta-\theta'}{2}\right)\, d\theta' \ .   \label{b34}
\eeqa
Now we can write $\{\bX^\lambda(\theta), \theta \}$ and the linear problem along the boundary direction $\theta$ becomes
\beq
 -\partial_\theta^2 \chi(\theta) - \frac{1}{2} \{ \bX^\lambda(\theta),\theta\} \, \chi=0\ ,
  \label{b35}
\eeq
 which should have anti-periodic solutions for any value of $\lambda$. Now we show that given this data it is possible to compute the area of the surface. 
 
\subsection{Expansion for the spectral parameter $\lambda\rightarrow 0$}

 It is useful to compute the behavior of the solution for $\lambda\rightarrow 0$. In this region it is convenient to introduce a new spectral parameter $y$ such that
\beq
 \lambda = - \frac{y^2}{4}\ .
\eeq
 The equations for $\psi_1,\psi_2$ are
\beqa
  \partial \psi_1 &=& -\half \partial\alpha \psi_1 - \frac{y^2}{4} e^\alpha \psi_2                     \label{b36} \\
  \partial \psi_2 &=& f e^{-\alpha} \psi_1 + \half \partial \alpha \psi_2                              \label{b37} \\
  \bar{\partial} \psi_1 &=& \half \bar{\partial} \alpha \psi_1 - e^{-\alpha} \bar{f} \psi_2            \label{b38} \\
  \bar{\partial} \psi_2 &=& -\frac{4}{y^2} e^{\alpha} \psi_1 - \half \bar{\partial}\alpha \psi_2    \ .   \label{b39}
\eeqa 
 Defining
\beq
  F= e^{\alpha} \frac{\psi_1}{\psi_2}\ ,
   \label{b40}
\eeq 
it is easy to find that
\beqa
 \partial F &=& -\frac{y^2}{4} e^{2\alpha} - f e^{-2\alpha} F^2                     \label{b41} \\
 \bar{\partial} F &=& 2 \bar{\partial} \alpha F - \bar{f} + \frac{4}{y^2} F^2  \ .     \label{b42}
\eeqa
 Now the expansion
\beq
 F = \pm \frac{y}{2} \sqrt{\bar{f}} + y^2 \left(\frac{1}{16}\bar{\partial} \ln \bar{f}-\frac{1}{4}\bar{\partial}\alpha\right) + \cO(y^3)
 \label{b43}\ ,
\eeq
follows. From here the corresponding expansion for $\psi_{1,2}$ is
\beqa
 \ln \psi_1 &=& -\half \alpha \mp \frac{2}{y} \int^{\bar{z}} \sqrt{\bar{f}} d\bar{z} + \frac{1}{4} + y \zeta_{11} + \cO(y^2)                  \label{b44} \\
 \ln \psi_2 &=& \half \alpha \mp \frac{2}{y} \int^{\bar{z}} \sqrt{\bar{f}} d\bar{z} - \frac{1}{4} \ln \bar{f} + y \zeta_{21} + \cO(y^2)   \ .     \label{b45}
\eeqa
 The coefficients $\zeta_{11}$ and $\zeta_{21}$ obey
\beqa
 \partial (\sqrt{\bar{f}} \zeta_{11}) &=& \mp \half e^{2\alpha}                  \label{b46} \\
 \partial (\sqrt{\bar{f}} \zeta_{21}) &=& \pm \half f\bar{f} e^{-2\alpha}  \ .      \label{b47}
\eeqa
 The main significance of these equations is that the corresponding right-hand sides are total derivatives, a fact that will be important when computing the area. 
For that purpose it is only necessary to know the coefficient $\zeta_{21}$ at the boundary. To obtain it, we need to solve the linear problem at the boundary for the function $\psi_2$. In analogy with eq.(\ref{b21}), we start by rewriting the functions $\psi_2$ as
 \beq
  \psi_2 = \sqrt{J_{12}} \chi  \ ,
   \label{b48}
 \eeq
 where $J_{12}$ behaves, near the boundary as
\beq
 J_{12} = -\frac{i}{\lambda} \frac{\sqrt{1-\xi}}{\xi} e^{-i\theta} \, + \cO(\xi)
\ , \label{b49}
\eeq
 and $\chi$ obeys the equation
\beq
  \partial_\theta^2\chi = V \chi = \frac{1}{y^2} V_{-1} + V_0 + y^2 V_1\ ,
 \label{b50}
\eeq
 with
\beq
 V_{-1} = -4\bar{f} e^{-2i\theta} , \ \ \ V_1 = \frac{1}{4} f e^{2i\theta} , \ \ \ V_0 = -\frac{1}{4} + 6\beta_2(\theta)\ ,
 \label{b51}
\eeq
similar to eq.(\ref{b24}).
 Writing
\beq
  \chi = e^S, \ \ \ \ S = \sum_{n=-1}^{\infty} y^n S_n\ ,
 \label{b52}
\eeq 
and using $\alpha = -\ln \xi + \cO(\xi^2)$ we find that
\beq
 \ln \psi_2 = \half \alpha - \half i\theta + \frac{1}{y} S_{-1} + S_0 + y S_1 +\ldots
 \label{b53}
\eeq 
Comparing with the previous result we should have
\beq
 S_{-1} = \mp 2 \int^{\bar{z}} \sqrt{\bar{f}} d\bar{z}, \ \ S_0 -\half i\theta = -\frac{1}{4} \ln \bar{f} , \ \ S_1 = \zeta_{21}\ .
 \label{b54}
\eeq
The coefficients $S_n$ can be found independently by solving the differential equation.   The first coefficient $S_{-1}$ turns out to be equal to
\beq
 S_{-1} = \pm \int_{0}^{\theta} \sqrt{V_{-1}(\theta')} d\theta'  = \pm 2 i \int_0^\theta e^{-i\theta'} \sqrt{\bar{f}} d\theta' = \mp 2\int^{\bar{z}} \sqrt{\bar{f}} d\bar{z}\ ,
 \label{b55}
\eeq  
 up to an arbitrary integration constant. The next two coefficients are then determined from
 \beqa
  S_0 &=& -\frac{1}{4} \ln V_{-1} = \half i\theta - \frac{1}{4}\ln \bar{f} +C_0                                                                      \label{b56} \\
  S'_1 &=& \pm \frac{1}{\sqrt{V_{-1}}} \left[V_0 + \frac{1}{4} \partial_{\theta}^2 V_{-1} - \frac{1}{16} (\partial_{\theta}V_{-1})^2\right]   \ ,       \label{b57}
 \eeqa
 where $C_0$ is an irrelevant constant. The values $S_{0}$ and $S_{-1}$ agree with the expectations and the value of $S_1$ determines the coefficient $\zeta_{21}=S_1$.
 The rest can be found recursively
 \beq
 S'_{n+1} = \frac{1}{2S'_{-1}} \left(-S''_n-\sum_{p=0}^n S'_pS'_{n-p}\right)\ ,
 \label{b58}
\eeq
 although they are not going to be needed in this paper. It is interesting to note that the periodicity condition
\beq
 \int_{0}^{2\pi} S'_{n+1} =0\ ,
\eeq
is non-trivial in terms of the right hand side of eq.(\ref{b58}). These conditions are equivalent to the vanishing of the holonomy and will be written later in a different way. 

\subsection{Computation of the area} 
 
 The area is defined, in principle, as
\beq
 A_\infty = 4 \int_D e^{2\alpha} d\sigma d\tau \ .
  \label{b59}
\eeq
 This quantity however is infinite. Introducing a regulator $\epsilon\rightarrow 0$ it is shown in the appendix that the area can be written as
\beq
 A_\infty = \frac{L}{\epsilon} - 2\pi -4 \int_{D} f\bar{f} e^{-2\alpha} d\sigma d\tau\ ,
 \label{b60}
\eeq
where $L$ is the length of the boundary curve.  Therefore the finite part of the area, and the one that is interesting for physical applications, is defined as
\beq
 A_f = -2\pi - 4 \int_D f\bar{f} e^{-2\alpha} d\sigma d\tau\ .
 \label{b61}
\eeq
 As a comment, this result implies that \cite{SARM}
\beq
 A_f \le -2\pi\ .
 \label{b62}
\eeq
 The equality is valid for the half-sphere whose boundary is a circle. Using eq.(\ref{b47}) we find
\beq
 A_f = -2\pi \mp 8 \int_D \partial (\sqrt{\bar{f}} \zeta_{21})\, d\sigma d\tau \ .
 \label{b63}
\eeq
 Choosing coordinates
 \beq
 z = \sigma+ i \tau = r e^{i\theta}\ ,
 \label{b64}
\eeq
 it follows that 
 \beq
  A_f = -2\pi \mp 4i \oint d\bar{z} \sqrt{\bar{f}} \zeta_{21}\ ,
  \label{b65}
 \eeq
 where the integral is over the unit circle in the $z$ plane. From the previous section we can use that
 \beq
  S'_{-1} = \pm 2i e^{-i\theta} \sqrt{\bar{f}}, \ \ \ S_1 = \zeta_{21}\ ,
   \label{b66}
 \eeq
  to write
\beq
 A_f = -2\pi \pm 2i \oint S'_{-1} S_1 d\theta = -2\pi \mp 2i \oint S_{-1} S'_1 d\theta \ ,
  \label{b67}
\eeq
 where we integrated by parts and use periodicity in $\theta$ to eliminate a boundary term. Using some algebra we obtain
\beqa
 A_f &=& -2\pi \mp i \oint \frac{S_{-1}}{S'_{-1}} \left(V_0 + \half \{ S_{-1},\theta\} \right) d\theta                                \label{b68} \\
     &=& -2\pi \pm \frac{i}{2} \oint \frac{S_{-1}}{S'_{-1}} \left(\mathrm{Re}\{\bX,\theta\} -  \{ S_{-1},\theta\} \right) d\theta  \ ,     \label{b69}
\eeqa
 where we used eqs.(\ref{b51},\ref{b28}), namely that $V_0=-\half \Re \{\bX,\theta\} $. This formula shows that if we indeed know the function $X(\theta)$, we can compute $f(\theta)$ from eqs.(\ref{b34},\ref{b27})
 and thus the area. As we have already mentioned a few times the function $X(\theta)$ is related ot the curve $X(s)$ by an unknown reparameterization $s(\theta)$. It is therefore useful to rewrite the formulas in 
terms of $X(\theta(s))$. Since for any function $F(\theta(s))$:
\beq
 \{ F, \theta\} = \{s,\theta\} + (\partial_\theta s)^2 \{ F,s \}  \ ,
 \label{b70}
\eeq
 we obtain
\beq
 A_f = -2\pi \pm \frac{i}{2} \oint \frac{\mathrm{Re} \{X,s\} - \{w,s\}}{\partial_s \ln w} ds \ ,
 \label{b71}
\eeq
where we defined the complex variable 
\beq
 w = \int^z \sqrt{f} dz \ ,
 \label{b72}
\eeq
 such that $S_{-1}=\pm 2 \bar{w}$. The sign in the equation should be chosen such that the condition (\ref{b62}) is satisfied, \ie\ $A_f\le -2\pi$. 
 In this form the expression for the area is manifestly reparameterization invariant (using eq.(\ref{b70}) to change parameterization $s\rightarrow s'$). 
If we consider $w(s)$ as a given (complex) function we can define $X(w)$ with the understanding that derivatives are taken as $\partial_w X(w)=\frac{\partial_s X}{\partial_s w}$. With such definition and using that $\Re\{X,s\} =\half(\{X,s\}+\{\bX,s\})$ and eq.(\ref{b70}) we find the following interesting expression for the area 
\beq
 A_f = -2\pi \pm \frac{i}{4} \oint \left[\{X,w\}+\{\bX,w\}\right] w\, dw\ ,
\eeq 
 indicating that, if we were to extend $X(w)$ to the interior of the contour $w(s)$, the area is related ot the double poles of $\{X(w),w\}$.
 
 Summarizing, since $X(s)$ is given, the problem of computing the area reduces to finding the complex contour $w(s)$. This is highly non-trivial and is the equivalent of finding  
the normal derivative given the value of the function in the Laplace problem. In that case the known solution is to use a dispersion relation such as eq.(\ref{b32}). Equivalently one can use the vanishing of an infinite set of conserved quantities as described in the Appendix.
In the present case we can rewrite the problem in an interesting way, as done in the next section,
but at present we cannot give a general solution. 

\section{The condition of vanishing charges}
\label{eqproblem}

 In the previous section we found that the area is completely determined if we have the contour $X(s)$ and the function $w(s)$. If we were given the function $X(\theta)$ in the conformal parameterization, then we could compute $w(s)$ by integrating $f(z)$ which is determined by the imaginary part of the Schwarzian derivative $\{\bX,\theta\}$.
 However, the function $\theta(s)$ that would determine $X(\theta)$ from $X(s)$ is unknown. Given $X(s)$ and eq.(\ref{b34}) we can write instead
\beq
  \{ \bX^\lambda, s\} = \{ \bX, s \} + \frac{i}{2} \left(\lambda+\frac{1}{\lambda} -2\right) \Im\{\bX,s\} -  \left(\lambda-\frac{1}{\lambda}\right) V_2(s) \ ,
   \label{c1}
\eeq
where we used the following property of the Schwarzian derivative
\beq
 \Re \{\bX,s\} = \{\theta,s\} + (\partial_s\theta)^2 \Re \{\bX,\theta\} , \ \ \ \Im \{\bX,s\} = (\partial_s\theta)^2 \Im \{ \bX,\theta\} \ ,
 \label{c2}
\eeq
and introduced the unknown function $V_2(s)$. From eqs.(\ref{c2}) and (\ref{b27}) we identify
\beq
V_2(s)+iV_1(s)=-2f(\theta)e^{2i\theta} (\partial_s\theta)^2 , \ \ \ \ V_1(s) =-\half \Im \{ \bX,s\} \ ,
\label{c3}
\eeq
and then\footnote{The square root should be defined such that $w(s)$ is continuous (and periodic).}
\beq
 w(s) = \int^s \sqrt{V_2(s) + i V_1(s)}\, ds \ .
 \label{c4}
\eeq
 Namely, the function $V_2(s)$ completely determines $w(s)$ and thus the area. It also gives the Schwarzian derivative as
 \beq
 \{\bX^\lambda(s),s\} = \Re\{\bX,s\} - \lambda (\partial_s w)^2 + \frac{1}{\lambda}(\partial_s\bar{w})^2 \ .
 \eeq
 
 To determine $V_2(s)$ we change variables $\theta\rightarrow s$ in eq.(\ref{b35})
obtaining 
\beq
 -\partial_s^2 \phi + V \phi =0, \ \ \ \ \ V(s)=- \half \{\bX^\lambda,s\} \ ,
 \label{c5}
\eeq
 where we replaced $\chi(\theta) = \sqrt{\partial_s\theta}\, \phi(s)$. For $\lambda=1$ the equation has two anti-periodic solutions 
\beq
 \phi_a(s) = \frac{1}{\sqrt{\partial_s \bX(s)}}, \ \ \ \ \phi_b = \frac{\bX(s)}{\sqrt{\partial_s \bX(s)}} \ .
 \label{c6}
\eeq
Notice that $\bX(s) = \phi_b(s)/\phi_a(s)$. For general values of $\lambda$ the solutions of such differential equation might not be anti-periodic. In particular take 
two solutions satisfying the boundary conditions
\beq
\phi_1(0)=1,\ \  \partial_s\phi_1(0)=0, \ \\ \  \phi_2(s)=0, \ \ \partial_s\phi_2(0)=1 \ .
\label{c7}
\eeq
 Since the potential $V(s)$ is periodic with periodicity $2\pi$, a shift in $s$ by $2\pi$ defines two new solutions \cite{Hill}
\beqa
 \tilde{\phi}_1(s) &=& \phi_1(s+2\pi) = A(\lambda) \phi_1(s) + B(\lambda) \phi_2(s),  \label{c8} \\
 \tilde{\phi}_2(s) &=& \phi_2(s+2\pi) = C(\lambda) \phi_1(s) + D(\lambda) \phi_2(s) \ .  \label{c9}
\eeqa
This defines the monodromy matrix as
\beq
 \Omega=\left(\begin{array}{cc} A(\lambda) & B(\lambda) \\ C(\lambda) & D(\lambda) \end{array}\right)\ .
 \label{c10}
\eeq
Since the Wronksian $W=\phi_1\phi'_2-\phi_2\phi'_1=1=AD-BC$ we have $\Omega\in SL(2\mathbb{C})$. The quasi-momentum $p(\lambda)$ is defined from the trace of the monodromy matrix as
\beq
 \tr\,{\Omega} = A(\lambda)+D(\lambda) = 2\cos(p(\lambda))\ ,
 \label{c11}
\eeq
and determines a set of conserved quantities. However, the monodromy matrix should be trivial. We know that the linear problem can be solved in the disk which has no non-trivial loops and therefore the solutions $\psi_{1,2}$ are periodic. Thus, the corresponding solutions $\chi$ are anti-periodic for any value of $\lambda$ and $p(\lambda)=\pi$.  Thus, the problem  of finding $V_2(s)$ reduces to the following problem: \\
{\bf Problem:}  Consider the equation
\beq
-\partial_s^2 \phi + V(\lambda,s) \phi(s) =0, \ \ \ \ \ V(\lambda,s)= V_0(s)  + \frac{i}{2} \left(\lambda+\frac{1}{\lambda} \right) V_1(s) + \frac{1}{2} \left(\lambda-\frac{1}{\lambda}\right) V_2(s)\ ,
\label{c12}
\eeq 
where $V_0(s):\mathbb{R}\rightarrow \mathbb{R}$ and $V_1(s):\mathbb{R}\rightarrow \mathbb{R}$ are known  periodic  functions of $s$ with period $2\pi$, determine the periodic function $V_2(s):\mathbb{R}\rightarrow \mathbb{R}$ such
that, for any value of $\lambda$,  all solutions $\phi(s)$ of the equation are antiperiodic in $s$, \ie\ $\phi(s+2\pi)=-\phi(s)$.  In our case $V_0(s) + i V_1(s) = -\half \{ \bX, s \}$ and the resulting $V_2(s)$ can be used in eqns.(\ref{c4}) and (\ref{b71}) to find the area. If one so prefers, defining the function $\psi=e^{\half is}\phi$ we can say that all solutions of the equation
\beq
 -\partial^2_s \psi + i \partial_s \psi + \left[V(\lambda,s)+\frac{1}{4}\right] \psi =0\ ,
\eeq
are periodic in $s=[0,2\pi]$, for any value of $\lambda$.

 In any case, this problem is equivalent to the following one: given a curve $\bX(s)$ in the complex plane (or Riemann sphere), determine a one complex parameter family of curves $\bX^\lambda(s)$ such that $\bX^{\lambda=1}(s)=\bX(s)$ and such that the Schwarzian derivative $\{\bX^{\lambda}(s),s\}$ 
is a meromorphic function of $\lambda$ with only a simple pole at $\lambda=0$ and a simple pole at infinity. That is
\beq
 \{\bX^\lambda(s),s\} = -2 \left[ V_0(s)  + \frac{i}{2} \left(\lambda+\frac{1}{\lambda} \right) V_1(s) + \frac{1}{2} \left(\lambda-\frac{1}{\lambda}\right) V_2(s) \right] \ ,
 \label{c13}
\eeq
 for some functions $V_{0,1,2}(s)$. Since $\bX^\lambda(s)=1$ is known, the functions $V_{0,1}$ are known, only $V_2(s)$ follows from solving this problem. 

The two problems are equivalent. Given a function $V_2(s)$ in the first problem, one can find two linearly independent solutions $\phi_1(s)$ and $\tilde{\phi}_1(s)$ that should be anti-periodic according to the statement of the problem. Taking
\beq
 \bX^\lambda(s)= \frac{\phi_1(s)}{\tilde{\phi}_1(s)}\ ,
 \label{c14}
\eeq
solves the second problem. Vice-versa, given a family $\bX^\lambda(s)$ that solves the second problem, $V_2$ follows. That all solutions of the linear problem are anti-periodic follows by simply exhibiting the following basis of solutions:
\beq
 \phi_a(s) = \frac{1}{\sqrt{\partial_s \bX^\lambda(s)}}, \ \ \ \ \phi_b(s) = \frac{\bX^\lambda(s)}{\sqrt{\partial_s \bX^\lambda(s)}}\ .
 \label{c15}
\eeq
If any of these two equivalent problems is solved, then the area of the minimal surface can be found. Unfortunately we were not able to find a generic analytical or numerical procedure to solve this problem. In the following we are going to give the solution of the case of small perturbations around the known circular Wilson loop and also a solution in terms of theta functions. 

\subsection{The R-function}

 Most of the paper is based on the Schwarzian derivative, a conformal invariant. In this subsection we just mention the possibility of defining another invariant. Given two linearly independent solutions of the boundary problem $\phi_1(s)$ and $\phi_2(s)$ normalized such that the Wronskian $W=\phi_1(s)\phi'_2(s) - \phi'_1(s) \phi_2(s)=1$ we define
\beq
 R^\lambda(s,s') = \phi_1(s) \phi_2(s') - \phi_1(s') \phi_2(s) \ .
  \label{c16}
\eeq 
 The main property of $R(s,s')$ is that it does not depend on which two solutions we choose. Namely if we consider two other (equally normalized) solutions:
\beq
  \tilde{\phi}_1 = a \phi_1 + b \phi_2, \ \ \  \tilde{\phi}_2 = c\phi_1+d\phi_2, \ \ \ ab-cd=1\ ,
   \label{c17}
\eeq
then the R-function defined with $\tilde{\phi}_{1,2}$ is the same in view of the condition $ab-cd=1$. Such condition is required for the Wronskian to be one. The function $R(s,s')$ is related to the local cross ratios \cite{KZ} defined as
\beq
 Y(s,s')=\frac{\partial_s \bX(s) \partial_{s'} \bX(s')}{(\bX(s)-\bX(s'))^2} = \frac{1}{R^2(s,s')}\ ,
\eeq
as can be seen by using $\bX(s)=\phi_1(s)/\phi_2(s)$ and the condition that the Wronskian is one. 
Other interesting properties are
\beq
 R^\lambda(s,s) =0, \ \  \left.\partial_s R^\lambda(s,s')\right|_{s'=s} = -1, \ \ \ R(s+2\pi,s')=-R(s,s')=R(s,s'+2\pi) \ ,
  \label{c18}
\eeq
 and the equation
\beq
 (\partial_s^2 -\partial^2_{s'}) R^\lambda(s,s') = (V(s)-V(s')) R^\lambda(s,s') \ ,
 \label{c19}
\eeq 
 where the potential $V(s)$ is the one in eq.(\ref{c5}). Equations of this type are studied for example in \cite{Marchenko} and could provide a good way to approach the problem but 
 we leave a more detailed study of the function $R(s,s')$ for future work. 

\section{Near circular Wilson loops}

 The near circular solution was originally studied by Semenoff and Young \cite{SY}, those results were extended to all loops in \cite{CHMS} by using a previous result from Drukker \cite{Drukker}. The integrable properties of those solutions were recently discussed in \cite{Cagnazzo}. Here we consider the near circular solutions as an illustration of the discussions in this paper. First we describe the solution in the usual approach of perturbing the equations of motion and then we show that the same
 results can be obtained, perhaps even more straight-forwardly using the methods of this paper. One caveat is that in our approach, the limit $\lambda\rightarrow 0$ is relevant but it does not commute with the small perturbation limit. For that reason we need to redo the computation of the area. Before going into the derivation 
let us summarize the results. The circular Wilson loop is a map from the unit disk parameterized by $z=re^{i\theta}$, $r\le 1$ into Poincare AdS such that the boundary $r=1$ maps to the circle $X=e^{i\theta}$. If we parameterize the fluctuations as
\beq
 X = e^{i\theta-\xi(\theta)}  \ ,
 \label{d1}
\eeq
 what we find in this paper is that we have to analytically continue the function $\xi(\theta)$ to a function $g(z)$ such that
\beq
 \xi(\theta) = 2 \Re[ g(z=e^{i\theta})] \ .
 \label{d2}
\eeq
 Having done that, the function $f(z)$ in the definition of the flat current, namely eq.(\ref{a10}) is given by
\beq
 f(z) = -\half ( z \partial^3 g + 3\partial^2 g) \ ,
 \label{d3}
\eeq 
and the area is given by
\beq
 A_f = -2\pi + i \oint d\theta \, g(\theta) \left(\partial_\theta^3 \bar{g}(\theta) + \partial_\theta \bar{g}(\theta)\right) \ ,
 \label{d4}
\eeq
where $\bar{g}(\theta)$ is the complex conjugate of $g(\theta)$. Let us see now how this is derived, first in the standard approach of perturbing the 
equations of motion and then with the method we are discussing, namely using the Schwarzian derivative of the contour. 

\subsection{Perturbing the equations of motion}
In the notation of this paper, the circular solution is given by
\beq
 \mathbb{A}_0 = \frac{1}{\sqrt{1-z\bz}} \left(\begin{array}{cc}1 & \bz \\ z & 1\end{array}\right), \ \ \ \ \mathbb{X}_0 = \mathbb{A}_0 \mathbb{A}_0^\dagger = \frac{1}{1-z\bz} \left(\begin{array}{cc}1+z\bz & 2\bz \\ 2z & 1+z\bz\end{array}\right) \ ,
 \label{d5}
\eeq 
 or equivalently, using $z=r e^{i\theta}$:
\beq
 Z = \frac{1-r^2}{1+r^2}, \ \ X = \frac{2r}{1+r^2} e^{i\theta}, \ \ \bar{X} = \frac{2r}{1+r^2} e^{-i\theta} \ ,
 \label{d6}
\eeq
 or, in embedding coordinates,
\beq
  X_1+iX_2 = \frac{2r}{1-r^2} e^{i\theta}, \ \ \ X_0 = \frac{1+r^2}{1-r^2}  \ \ \ \ \ X_3 =0 \ ,
  \label{d7}
\eeq
The flat current is
\beq
 J = \frac{1}{1-z\bar{z}}\left(\begin{array}{cc}
 -\half \bar{z} & 0 \\ \lambda & \half \bar{z}
 \end{array}\right) , \ \ \ \ \
 \bar{J} = \frac{1}{1-z\bar{z}}\left(\begin{array}{cc}
  \half z & \frac{1}{\lambda} \\ 0 & -\half z
  \end{array}\right) \ ,
\label{d8}
\eeq
so that
\beq
 e^{\alpha} = \frac{1}{1-z\bar{z}}, \ \ \ f(z) =0\ .
 \label{d9}
\eeq
 If the function $f(z)$ vanishes, we obtain the circular solution, therefore we need to consider a first order fluctuation such that $f(z)\neq 0$. By looking
 at eq.(\ref{b3}) we see that the variation of $\alpha$ is second order and therefore it can be ignored. Thus, the variation of the flat current is simply
\beq
 \delta J = (1-z\bar{z}) f(z) \sigma_+, \ \ \ \delta \bar{J} = - (1-z\bar{z}) \bar{f} \sigma_-\ .
 \label{d10}
\eeq
Notice that 
\beq
 \tr (\delta J \, \delta\bar{J}) =  (1-z\bar{z})^2 f\bar{f} = - e^{-2\alpha}f\bar{f}\ ,
 \label{d11}
\eeq
 implying, from eq.(\ref{b61}) that
\beq
 A_f = -2\pi + 4 \int \tr(\delta J \, \delta\bar{J}) d\sigma d\tau\ .
 \label{d12}
\eeq 
 A fluctuation in $\mathbb{A}$ can be parameterized as
 \beq
 \mathbb{A}  = e^{H} \mathbb{A}_0 \simeq \mathbb{A}_0 + H \mathbb{A}_0 \ ,
 \label{d13}
 \eeq
for a traceless matrix $H$ that should obey
\beq
\partial H = \mathbb{A}_0 \delta J\mathbb{A}_0^{-1} , \ \ \ \  \bar{\partial} H = \mathbb{A}_0 \delta \bar{J}\mathbb{A}_0^{-1}\ ,
\label{d14}
\eeq
namely
\beq
 \partial H = f(z) \left(\begin{array}{cc}
  -\lambda z & 1 \\ -\lambda^2 z^2 & \lambda z
  \end{array}\right) , \ \ \ 
\bar{\partial} H = - \bar{f} \left(\begin{array}{cc}
 -\frac{1}{\lambda} \bar{z} & -\frac{1}{\lambda^2} \bar{z}^2 \\ 1 & -\frac{1}{\lambda} \bar{z}
 \end{array}\right) \ .
 \label{d15}
\eeq
 The solution follows by simple integration and is more conveniently written in terms of a holomorphic function 
\beq
 g(z) = -z \int^z f(z') dz' + 2 \int^z z' f(z') dz' - \frac{1}{z} \int^z z'{}^2f(z')dz' \ ,
 \label{d16}
\eeq
so that
\beq
f(z) = -\half(z\partial^3 g+ 3\partial^2 g) \ .
\label{d17}
\eeq
Here it might not be clear why we define $g(z)$ this way but in the next subsection it appears quite naturally.  Now we have
\beq
H = H(z,\lambda) - \left(H(z,-\frac{1}{\bar{\lambda}})\right)^\dagger \ ,
\label{d18}
\eeq
with
\beq
H(z,\lambda) = \frac{\lambda}{2}(z^2\partial^2 g + z\partial g - g) \sigma_z -(\partial g+\half z \partial^2 g) \sigma_+ + \half \lambda^2 z^3 \partial^2 g \sigma_- \ .
\label{d19}
\eeq
 Given $H$ we can reconstruct the matrix $\mathbb{A}$ and from there the shape of the boundary contour as
 \beqa
 \bX &=& \left.\frac{\mathbb{A}_{11}}{\mathbb{A}_{21}}\right|_{r=1} \simeq \left.\frac{1}{\lambda} e^{-i\theta} \left(1+H_{11}-H_{22}+\lambda zH_{12} - \frac{1}{\lambda z} H_{21}\right)\right|_{r=1}   \label{d20}\\
   &\simeq& \frac{1}{\lambda} e^{-i\theta-\lambda g(z) - \frac{1}{\lambda} \bar{g}(\bar{z})}    \ .   \label{d21}
\eeqa
Taking $\lambda=1$ as the original contour and matching with eq.(\ref{d1}) we find that $g(z)$ is an analytic function in the disk whose boundary value is determined by the fluctuation $\xi(\theta)$  as
\beq
 \xi(\theta) = 2 \Re[ g(z=e^{i\theta})] \ .
 \label{d22}
\eeq
 This completely fixes the function $g(z)$. 
 
 The Area can be computed from eq.(\ref{d12}) by noticing from eqs. (\ref{d14}) and (\ref{d18}) that
\beq
 \tr(\delta J \, \delta\bar{J}) = \tr(\partial H \bar{\partial} H) = - \partial\left\{\tr\left[ H(z,\lambda)  \bar{\partial}\left(H(z,-\frac{1}{\bar{\lambda}})\right)^\dagger \right]\right\} \ .
 \label{d23}
\eeq
 Integrating by parts and using the value for $H$ from eq.(\ref{d19}) we find
\beqa
 A_f &=& -2\pi - 2i \oint d\theta \tr(H(z,\lambda) \partial_\theta\left(H(z,-\frac{1}{\bar{\lambda}})\right)^\dagger )  \label{d24a}\\ 
      &=& -2\pi + i \oint d\theta \, g(\theta) \left(\partial_\theta^3 \bar{g}(\theta) + \partial_\theta \bar{g}(\theta)\right)  \ .  \label{d24b}
\eeqa
The surface itself can be obtained by replacing $H$ in eq.(\ref{d13}) and then using $\mathbb{X} =\mathbb{A} \mathbb{A}^\dagger$. In this way it follows that, in Poincare
coordinates, the perturbative solution is
\beqa
 \delta Z &=& \frac{1-z\bar{z}}{1+z\bar{z}}\, \left[ g(z) +\bar{g}(\bar{z}) + \frac{1-z\bar{z}}{1+z\bar{z}} (z\partial g(z)+\bar{z}\bar{\partial} \bar{g}(\bar{z}) \right] \label{d24c}\\
 \delta X &=& \frac{2z}{1+z\bar{z}} \left[ g(z) +\bar{g}(\bar{z}) +\frac{1-z\bar{z}}{1+z\bar{z}} (z\partial g(z)+\bar{z}\bar{\partial} \bar{g}(\bar{z}) )\right] \ . \label{d24d}
\eeqa

\subsection{Derivation using the Schwarzian derivative}

 The method described in this paper is particularly simple for this case because the fluctuations do not affect $\alpha$ meaning that the world-sheet metric remains
 conformal and therefore the parameterization $X(\theta)=e^{i\theta-\xi(\theta)}$ is already conformal!, \ie\ we do not need to find the reparameterization $s(\theta)$. 
 The Schwarzian derivative of the contour $\bX= e^{-i\theta-\xi(\theta)}$ is, at first order in $\xi$:
 \beq
 \{ \bX ,\theta\} = \half - i (\partial_\theta^3\xi+\partial_\theta\xi)\ ,
 \label{d25}
\eeq
and thus
\beqa
 \Re\, \{\bX,\theta\} &=& \half\ \ \ \ \Rightarrow \ \ \ \beta_2(\theta)=0 \label{d25b}\\
 \Im\, \{\bX,\theta\} &=& -(\partial_\theta^3\xi+\partial_\theta\xi) = -2 i (f(\theta) e^{2i\theta} - \bar{f}(\theta) e^{-2i\theta} ) \ ,  \label{d25c}
\eeqa
where we used eq.(\ref{b27}) with $\lambda=1$ or, equivalently, eqs.(\ref{b28}, \ref{b29}). Since $\xi(\theta)$ is periodic we can write it as
\beq
 \xi(\theta) = \xi_0 + \sum_{n\ge1}\left(\xi_n e^{in\theta} + \bar{\xi}_n e^{-in\theta}\right) \ . \label{d25d}
\eeq   
 On the other hand $f(z)$ is analytic in the unit circle and then
\beq
 f(z) = \sum_{n\ge0} f_n z^n  \ \ \ \ \Rightarrow \ \ \ \ f(\theta) = \sum_{n\ge 0} f_n e^{in\theta} \ .  \label{d25e}
\eeq
 It follows that
\beq
 f_{n-2} = \half n (1-n^2) \xi_n \ .  \label{d25f}
\eeq
 Equivalently, if we define the analytic function
\beq
 g(z) = \half \xi_0 + \sum_{n\ge1} \xi_n z^n \ ,  \label{d25g}
\eeq
 and use the relation (\ref{d25f}), we obtain
\beq
 f(z) = -\half (z\partial^3g+2 \partial^2 g) \ , \label{d25h}
\eeq
that justifies this somewhat curious expression we introduced in eq.(\ref{d3}).
This completes the calculation of the analytic function  $f(z)$, the only  data we needed to compute the area:
\beq
 A_f = -2\pi - 4 \int_D f\bar{f}\, e^{-2\alpha} d\sigma d\tau
  = -2\pi - 4 \int_D f\bar{f}\, (1-z\bar{z})^2 d\sigma d\tau \ , \label{d25i}
\eeq
where we used eqns.(\ref{b61}) and (\ref{d9}). Previously we used the limit $\lambda\rightarrow 0$ to show that the integrand is a total derivative. This limit is tricky here since $\frac{1}{\lambda}\rightarrow\infty$ in the Schwarzian derivative (\ref{b27}) violating the condition of small perturbation. However the integrand can be shown to be a total derivative by simple inspection:
\beqa
 \partial F &=& f\bar{f}\, (1-z\bar{z})^2 = -\half \bar{f} (z\partial^3g+2 \partial^2 g ) \, (1-z\bar{z})^2   \label{d25j}\\
  F&=& -\bar{f} \left[\half z (1-z\bar{z})^2 \partial^2 g + (1-z\bar{z}) \partial g + \bar{z} g\right] \ . \label{d25k}
\eeqa
 In this way the area simplifies to
\beq
  A_f = -2\pi -2 \oint d\theta\, e^{-i\theta} F(r=1,\theta) \ .  \label{d25l}
\eeq 
 But
\beq
 F(r=1,\theta) = -e^{-i\theta}\,\bar{f}(\theta) g(\theta) \ .   \label{d25m}
\eeq
From eq.(\ref{d25h}) and expanding the derivatives it follows that
\beqa
 A_f &=& -2\pi + \oint d\theta\, e^{-2i\theta}\, \bar{f}(\theta) g(\theta) \\
     &=& -2\pi - \oint d\theta e^{-2i\theta} g(\theta) \left.\left[\bar{z} \bar{\partial}^3 \bar{g} + 3 \bar{\partial}^2 \bar{g}\right]\right|_{r=1} \\
     &=& -2\pi + i \oint d\theta g(\theta)\left(\partial_\theta\bar{g} +\partial_\theta^3\bar{g}(\theta)\right) \ ,
\eeqa
namely the same expression derived in eq.(\ref{d24b}). To complete this subsection let us mention that the boundary linear problem is
\beq
 - \partial_\theta^2 \chi  + V \chi = 0 \ ,
 \label{d26}
\eeq
where
\beq
 V = -\half \{ \bX^\lambda,\theta\} = -\frac{1}{4} +\frac{i\lambda}{2} (\partial^3_\theta g + \partial_\theta g) + \frac{i}{2\lambda} (\partial^3_\theta \bar{g} + \partial_\theta \bar{g}) \ ,
 \label{d27}
\eeq
 with two anti-periodic solutions 
\beqa
 \chi_1  &=& e^{-\half i\theta} \left[1+\half\lambda(g-i\partial_\theta g)+\frac{1}{2\lambda}(\bar{g}-i\partial_\theta\bar{g}) \right]    \label{d28}\\
 \chi_2 &=& e^{\half i\theta} \left[1-\half\lambda(g+i\partial_\theta g)-\frac{1}{2\lambda}(\bar{g}+i\partial_\theta\bar{g})\right]   \ ,    \label{d29}
\eeqa
at first order in the perturbation. It should be noted that these expressions are not valid in the limits $\lambda\rightarrow 0,\infty$ since the corrections would not be small. 

\section{Solution in terms of theta functions}

 In \cite{IKZ,KZ} an infinite parameter family of minimal area surfaces was obtained analytically in terms of theta functions. Those solutions provide an infinite parameter family of solutions to the problem described in section \ref{eqproblem}. We are going to write here those solutions. Making this section self-contained would make it too long and take it out of the scope of this paper. For that reason we refer the reader to \cite{KZ} for definitions, notation and the identities necessary to prove that these are indeed solutions of the problem in section \ref{eqproblem}. General references on theta functions are, for example, \cite{ThF,FK} and their application to integrable systems can be found \eg\ in \cite{BBook}.
 
 In our case, first we introduce two theta functions $\theta$, $\hat{\theta}$ associated with a hyperelliptic Riemann surface and such that they differ by an odd half-period. Then a vector $\zeta(s)\in \mathbb{C}^g$ is defined as $\zeta(s)=2\omega_3 z(s) + 2\omega_1 \bar{z}(s)$ where $\omega_{1,3}$ are constant vectors and $z(s)$ is a closed curve in the complex plane such that $\hat{\theta}(\zeta(s))=0$ for any $s$. The potential is then given by
 \beqa
  V(s) &=& -\half \{z,s\} + \lambda (\partial_sz)^2 -\frac{1}{\lambda} (\partial_s \bar{z})^2 \\
  &&\ \  - 4(\partial_s z)^2 \left[2D_3\ln\theta(\zeta(s)) - \frac{D_3^3\hat{\theta(0)}}{D_3\hat{\theta}(0)} + \frac{D_3^2\theta(0)}{\theta(0)}\right] \ ,
  \label{e1}
 \eeqa
 where $D_3$ indicate derivative in the direction of the vector $\omega_3$. For any value of $\lambda$ there are two anti-periodic solutions to
\beq
 - \partial_s^2 \phi (s) + V(s) \phi(s) =0 \ ,
  \label{e2}
\eeq
given by 
\beqa
 \phi_a &=& \frac{1}{\sqrt{\partial_s z}}\,  \frac{\hat{\theta}(\zeta(s)-\int_1^4)}{\theta(\zeta(s))}\ e^{-\mu z-\nu \bar{z}}    \label{e3}\\
 \phi_b &=& \frac{1}{\sqrt{\partial_s z}}\,  \frac{\hat{\theta}(\zeta(s)+\int_1^4)}{\theta(\zeta(s))}\ e^{\mu z +\nu \bar{z}}  \ ,   \label{e4}
\eeqa 
where $4$ denotes a point in the hyperelliptic Riemann surface whose projection on the complex plane is $\lambda$. The constants $\mu$, $\nu$ are given by
\beq
 \mu=-2D_3\ln\theta(\int_1^4), \ \ \ \ \nu=-2D_1\ln\hat{\theta}(\int_1^4) \ .
  \label{e5}
\eeq
 It is clear then that the two solutions correspond to choosing $p_4$ as each of the two points on the Riemann surface that project to the save value of $\lambda$. We found these solution by using the already known results of \cite{KZ}.
 It should be interesting to use the techniques of \cite{BBook} to directly solve the problem. 
 
\section{Conclusions}

 In this paper we have studied the problem of finding the area of the minimal surface bounded by a given contour in the boundary of three dimensional hyperbolic space. We were able to give a formula for the area that depended on finding the correct parameterization for the contour, in close analogy to the case in flat space. To determine the correct parametrization we recast the problem as the one of finding a potential $V_2(s)$ that satisfies a curious property equivalent to the vanishing of the monodromy, or equivalently of the conserved charges. In the case of flat space the correct parameterization can in principle be found by minimizing a certain functional. 
 In our case, the problem of finding $V_2(s)$ seems more challenging. In fact, we do not know of a general analytic or numerical procedure to solve it. It seems that the problem can be treated at least numerically but we leave that for future work. It would be interesting to relate this problem to the TBA equations appearing in an alternative approach based on taking the limit of light-like Wilson loops and developed in \cite{Toledo}.
 
 More broadly, and speaking generally, the method we discussed can be thought as converting a boundary problem into an initial value problem 
for which integrability methods are more suited. In analogy with the Laplace equation, the vanishing of the conserved charges is the integrability  equivalent of the holomorphicity condition that relates the boundary value of the function with the value of the normal derivative. In string theory language, it determines the semi-classical state of the string. It is reasonable to speculate that the same idea might be extended to the quantum case and used to determine a boundary state for the string. 
 
 \section{Acknowledgments}
 
  The author wants to thank N. Drukker, J. Maldacena, J. Toledo, 
  A. Sever, A. Tseytlin, P. Vieira and S. Ziama for discussions on this topic. In addition he also wants to thank Imperial College (London), KITP, Santa Barbara and Perimeter Institute (Waterloo) for hospitality while this work was being done. 
  This work  was supported in part by NSF through a CAREER Award PHY-0952630 and by DOE through grant \protect{DE-SC0007884}.

\appendix

\section{The Plateau problem in flat space}

   The Plateau problem is to prove the existence of a minimal surface bounded by a given contour in $\mathbb{R}^n$. It was solved in the 1930s by Jesse Douglas \cite{JD} by 
writing a certain functional over the possible reparameterizations of the contour and showing that it always has a minimum and that such minimum defines the minimal surface. We are not concerned here with the details of the proof but instead with the techniques he used because they have some parallel with what we tried to do in this paper. In fact the usefulness of that construction for the AdS case was already pointed out in \cite{Ambjorn}.

Using the same notation than in the main part of the paper,  the surface is given by a map of the unit disk $|z|\le 1$ with $z=\sigma+i \tau=r e^{i\theta}$ into $\mathbb{R}^n$ through functions $X_{i=1\ldots n}(z,\bar{z})$.
 If the parameterization is conformal, the area is given by
\beq
 A = \half \int \left[ (\partial_\sigma X_i)^2 + (\partial_\tau X_i)^2 \right] d\sigma d\tau \ .
\eeq
 The equations of motion are
\beq
  \partial\bar{\partial} X_i=0\ ,
  \label{eomr}
\eeq 
solved by 
\beq
 X_i = \half \left[ g_i(z) + \bar{g}_i(\bar{z}) \right] \ ,
\eeq
 where $g_i(z)$ are holomorphic functions that can be determined from the boundary value $ \Re\, g(e^{i\theta}) =x_i(\theta)$. At the boundary we are going to write 
\beq
 g_i(e^{i\theta}) = x_i(\theta) + i \xi_i(\theta) \ ,
\eeq
where $\xi_i(\theta)$ is the imaginary part that can be determined by the dispersion relation:
\beq
 \xi_i(\theta_0) = -\frac{1}{2\pi} \strokedint x_i(\theta) \cotan\left(\frac{\theta-\theta_0}{2}\right) \ d\theta  \ .
 \label{r1}
\eeq 
Before continuing let us just mention that the dispersion relation is equivalent to the vanishing of an infinite set of conserved quantities given by
\beq
 q_n = \oint dz\, z^n \partial\phi, \ \ \ n\in \mathbb{Z}_{\ge0} \ .
\eeq
The reason we call the $q_n$ conserved quantities is that they are given by the holonomy of the conserved currents $j_n=z^2\partial\phi$, namely the $j_n$ satisfy 
$\bar{\partial} j_n=0$. Going back to the condition $q_n=0$, it relates $x_i(\theta)$ and $\xi_i(\theta)$ through
\beq
 q_n = \frac{i}{2} \oint d\theta \, e^{in\theta} (\partial_\theta\xi_i-i\partial_\theta x_i) =0 \ ,
\eeq
which after expanding $x_i(\theta)$ and $\xi(\theta)$ in Fourier modes:
\beqa
 x_i(\theta) &=& x_{i0} +\half \sum_{n=1}^{\infty} (x_{in} e^{in\theta} + \bar{x}_{in} e^{-in\theta})  \\
  \xi_i(\theta) &=& \xi_{i0} +\half \sum_{n=1}^{\infty} (\xi_{in} e^{in\theta} + \bar{\xi}_{in} e^{-in\theta}) \ ,
\eeqa
implies
\beq
 \bar{\xi}_{in} = i \bar{x}_{in}, \ \ \ \forall n<0 \ ,
\eeq
namely
\beq
 \xi_i(\theta) = \xi_{i0} - \frac{i}{2} \sum_{n=1}^{\infty} ( x_{in} e^{in\theta} - \bar{x}_{in} e^{-in\theta}) \ .
\eeq
Therefore, the condition $q_n=0$, $\forall n\ge0$ is equivalent to the statement that $x_i(\theta)$ and $\xi_i(\theta)$ are the real and imaginary part of the boundary value of the holomorphic function
\beq
g_i(z) = x_{i0} +i \xi_{i0} + \sum_{n\ge 1} x_{in} z^n  \ ,
\eeq
or equivalently to the dispersion relation (\ref{r1}). After this brief detour, let us go back to the main problem. 

 The equations of motion (\ref{eomr}) should be supplemented by the conformal constraint that reads 
\beq
(\partial g_i )^2 =0 \ .
\eeq
Since $(\partial g_i )^2$ is a holomorphic function it is enough to impose that its imaginary part vanishes at the boundary of the disk to ensure that it vanishes everywhere. Namely, we only need
\beq
 \partial_\theta x_i \partial_\theta \xi_i=0 \ .
\eeq
Now we can compute the area by simple integration by parts obtaining
\beq
A = \half \oint X_i \partial_r X_i d\theta = \half \oint x_i(\theta) \partial_\theta \xi_i(\theta) \, d\theta \ .
\eeq
We have
\beq
 \partial_\theta \xi_i(\theta_0) = -\frac{1}{4\pi} \strokedint \frac{x_i(\theta)-x_i(\theta_0)}{\sin^2\left(\frac{\theta-\theta_0}{2}\right)} \ .
\eeq
From here, the area, after symmetrizing the expression is given by
\beq
 A = \frac{1}{16\pi} \oint\oint \frac{(x_i(\theta)-x_i(\theta_0))^2}{\sin^2\left(\frac{\theta-\theta_0}{2}\right)}\ d\theta\, d\theta_0 \ .
 \label{h23}
\eeq
It seems that the problem of computing the area given the contour $x_i(\theta)$ is solved but, of course, the issue is the one that we mentioned before, we are only given $x_i(s)$ in some parameterization and we need to allow for an unknown  reparameterization $\theta(s)$ so that we can choose conformal coordinates on the disk. How do we choose $\theta(s)$?. If we take a wrong $\theta(s)$ the functions $x_i(\theta)$ are still defined and we can always analytically continue them to compute $\xi_i(\theta)$ and also compute the (wrong) area using eq.(\ref{h23}). The mistake will only appear if we check the conformal constraint, namely 
\beq
  0= \partial_\theta x_i \partial_\theta \xi_i = -\frac{1}{4\pi}\strokedint \frac{\partial_\theta x_i(\theta)\left[x_i(\theta)-x_i(\theta_0)\right]}{\sin^2\left(\frac{\theta-\theta_0}{2}\right)}  \ ,
  \label{h24}
\eeq
will not actually vanish for the wrong parameterization. As pointed out by Jesse Douglas, from all possible reparameterizations $\theta(s)$ the one that minimizes the formula (\ref{h23}) is the same one that satisfies the conformal constraint. Indeed, if the parameterization changes by $\theta(s)=\theta(s)+\delta\theta(s)$ the variation of $x_i(\theta)$ is
\beq
\delta x_i = \partial_\theta x_i \delta \theta \ .
\eeq
 Performing such variation in eq.(\ref{h23}) thought as a functional of the parameterization $\theta(s)$ shows that the condition for the variation to vanish is precisely the conformal constraint written as in eq.(\ref{h24}).

Having summarized the flat space case, we just want to take away two simple ideas. The area is determined by the contour and the normal derivatives of the functions $X_i(r,\theta)$ at the boundary. The latter can be obtained from a dispersion relation if the correct parameterization $\theta(s)$ is known. In this case there is a very beautiful result that the correct parameterization minimizes a functional whose minimum value can be identified with the area.
In our case the unknown parametrization was rewritten in terms of the potential $V_2(s)$ and determined from the condition that all charges vanish.

\section{Definition of the renormalized area}

In this appendix we derive the formula for the finite part of the area. This derivation can be found in \cite{IKZ} but we include it here for completeness since computing the area is the main purpose of this paper. 
 The area is defined naively as 
\beq
 A_\infty = 4 \int_D e^{2\alpha} d\sigma d\tau  \ ,
  \label{f1}
\eeq
but, as already mentioned this definition fails since the integral diverges near the boundary. The divergence is regulated by taking a contour of fixed $Z=\epsilon$ and expanding the area as
\beq
 A_\epsilon = 4 \int_{Z\ge\epsilon} e^{2\alpha} d\sigma d\tau  = \frac{L}{\epsilon} + A_f + \cO(\epsilon) \ ,
 \label{f2}
\eeq
 the divergent term is known to be given by the length of the contour and the finite piece $A_f$ is the one we are after. Using the equation of motion for $\alpha$ we find
\beq
 A_\epsilon = \oint (\nabla\alpha.\hat{n})\, d\ell - 4\int f\bar{f} e^{-2\alpha} d\sigma d\tau \ ,
 \label{f3}
\eeq 
where $\hat{n}$ is a unit vector normal to the contour $Z=\epsilon$, namely
\beq
 \hat{n} = - \frac{\nabla Z}{|\nabla Z|} \ .
 \label{f4}
\eeq
The functions $X$, $\bar{X}$ and $Z$ are regular in the disk including the boundary. The equations of motion imply
\beq
 \nabla X \nabla Z = \half Z \nabla^2 X \ ,
 \label{f5}
\eeq  
 namely that $\nabla X$ and $\nabla Z$ are perpendicular at $Z=0$ and also $\nabla X \nabla Z \sim \epsilon$ when $Z=\epsilon$. Furthermore, the equation of motion for $Z$ implies
\beq
 (\nabla Z)^2 - |\nabla X|^2  =  Z \nabla^2 Z \ .
 \label{f6}
\eeq
 Finally, near the boundary, $Z$ behaves as
\beq
 Z = e^{-\alpha}\, h \ ,
 \label{f7}
\eeq 
where $h$ is a non-vanishing positive function. Using that the length of the contour is given by
\beq
 L = \oint |\nabla X| d\ell \ ,
 \label{f8}
\eeq
 it follows that
\beq
 A_\epsilon = \frac{L}{\epsilon} +\half \oint \frac{\nabla^2 e^{-\alpha}}{|\nabla e^{-\alpha}|} d\ell - 4 \int f\bar{f} e^{-2\alpha} d\sigma d\tau \ ,
 \label{f9}
\eeq
 substituting the expansion (see eq.(\ref{b5}))
\beq
 e^{-\alpha} = \xi + \cO(\xi^3) \ ,
 \label{f10}
\eeq
it follows that
\beq
A_f = -2\pi - 4 \int f\bar{f} e^{-2\alpha} d\sigma d\tau \ ,
\label{f11}
\eeq
as used in the main text.

\section{Wavy Wilson line}

 In \cite{SY}, a Wilson loop with the shape
\beq
 X = s + i \zeta_1(s)  \ ,
 \label{g1}
\eeq
 was considered in the limit $|\dot{\zeta}| \ll 1$ as a perturbation of the straight line $X=s$. Here $\dot{\zeta}=\partial_s\zeta$. The area was found to be given by
\beq
 \delta A_f =  -\frac{1}{4\pi} \int_{-\infty}^{\infty} ds\, ds' \, \frac{\left(\dot{\zeta}_1(s)-\dot{\zeta}_1(s')\right)^2}{(s-s')^2} \ .
 \label{g2}
\eeq
 Consider now an analytic function $\zeta(w=s+i\tau)$ in the upper half plane $\Im(w)>0$ such that its real part, on the real axis $\tau=0$ equals $\zeta_1(s)$.
Let us denote the imaginary part on the real axis as $\zeta_2(s)$, namely $\zeta(s)=\zeta_1(s)+i\zeta_2(s)$. By a standard dispersion relation we have
\beq
 \zeta_2(s') = -\frac{1}{\pi} \strokedint \frac{\zeta_1(s)}{s-s'}\, ds \ ,
 \label{g3}
\eeq
and further
\beq
 \partial_{s'} \strokedint \frac{\dot{\zeta}_1(s)}{s-s'}\, ds =  \strokedint \frac{\dot{\zeta}_1(s)-\dot{\zeta}_1(s')}{(s-s')^2}\, ds \ .
 \label{g4}
\eeq
 Integrating by parts and using some algebra we then find
\beq
  \delta A_f = \half \int_{-\infty}^{\infty} \ddot{\zeta}_1(s) \dot{\zeta}_1(s)\, ds \ .
  \label{g5}
\eeq 
 To match with the formula (\ref{d24b}) in the main text we consider the near circular Wilson loop
\beq
 X = e^{i\theta+2g_1(\theta)} \ ,
 \label{g6}
\eeq
and do a conformal transformation to the wave Wilson line by doing
\beq
 \tilde{X} = -i \frac{X+1}{X-1} = -\cotan\frac{\theta}{2} - i \frac{1}{\sin^2\frac{\theta}{2}} g_1(\theta) \ .
 \label{g7}
\eeq
 We identify then
\beq
  s= - \cotan \frac{\theta}{2} , \ \ \ \ \zeta_1(s) = - \frac{1}{\sin^2\frac{\theta}{2}} g_1(\theta) \ .
  \label{g8}
\eeq 
Using the same map for the world-sheet, namely 
\beq
 w = -i \frac{z+1}{z-1} \ ,
 \label{g9}
\eeq
we find that the interior of the unit disk $|z|<1$ maps to the upper half plane $\Im(w)>0$ and therefore we identify the analytic function $\zeta(w)$ as
\beq
 \zeta(w) = - (1+w^2) g(z(w)) \ ,
 \label{g10}
\eeq
and thus
\beq
 \zeta_2(s) = - \frac{1}{\sin^2\frac{\theta}{2}} g_2(\theta) \ .
 \label{g11}
\eeq
 Replacing in eq.(\ref{g5}) and expanding it follows that
\beq
 \delta A_f = -2 \int_0^{2\pi} d\theta\, g_2(g_1'+g_1''') \ ,
 \label{g12}
\eeq
 in agreement with eq.(\ref{d24b}).


\begin{thebibliography}{99}        
\bibitem{malda}
J.~Maldacena,
``The large $N$ limit of superconformal field theories and supergravity,''
Adv.\ Theor.\ Math.\ Phys.\  {\bf 2}, 231 (1998)
[Int.\ J.\ Theor.\ Phys.\  {\bf 38}, 1113 (1998)],
{\tt hep-th/9711200}, \\
S.~S.~Gubser, I.~R.~Klebanov and A.~M.~Polyakov,
``Gauge theory correlators from non-critical string theory,''
Phys.\ Lett.\ B {\bf 428}, 105 (1998)
[arXiv:hep-th/9802109], \\
E.~Witten,
``Anti-de Sitter space and holography,''
Adv.\ Theor.\ Math.\ Phys.\  {\bf 2}, 253 (1998)
[arXiv:hep-th/9802150]. 

\bibitem{MRY}
  J.~M.~Maldacena,
  ``Wilson loops in large N field theories,''
  Phys.\ Rev.\ Lett.\  {\bf 80}, 4859 (1998)
  [arXiv:hep-th/9803002], \\
  S.~J.~Rey and J.~T.~Yee,
  ``Macroscopic strings as heavy quarks in large N gauge theory and anti-de
  Sitter supergravity,''
  Eur.\ Phys.\ J.\  C {\bf 22}, 379 (2001)
  [arXiv:hep-th/9803001].

\bibitem{cWL}
  D.~E.~Berenstein, R.~Corrado, W.~Fischler and J.~M.~Maldacena,
  ``The Operator product expansion for Wilson loops and surfaces in the large N
  limit,''
  Phys.\ Rev.\  D {\bf 59}, 105023 (1999)
  [arXiv:hep-th/9809188], \\
  D.~J.~Gross and H.~Ooguri,
   ``Aspects of large N gauge theory dynamics as seen by string theory,''
   Phys.\ Rev.\  D {\bf 58}, 106002 (1998)
   [arXiv:hep-th/9805129], \\
  J.~K.~Erickson, G.~W.~Semenoff and K.~Zarembo,
  ``Wilson loops in N = 4 supersymmetric Yang-Mills theory,''
  Nucl.\ Phys.\  B {\bf 582}, 155 (2000)
  [arXiv:hep-th/0003055], \\
  N.~Drukker and D.~J.~Gross,
  ``An exact prediction of N = 4 SUSYM theory for string theory,''
  J.\ Math.\ Phys.\  {\bf 42}, 2896 (2001)
  [arXiv:hep-th/0010274], \\
  V.~Pestun,
  ``Localization of gauge theory on a four-sphere and supersymmetric Wilson
  loops,''
  arXiv:0712.2824 [hep-th], \\
    M.~Kruczenski and A.~Tirziu,
    ``Matching the circular Wilson loop with dual open string solution at 1-loop
    in strong coupling,''
    JHEP {\bf 0805}, 064 (2008)
    [arXiv:0803.0315 [hep-th]], \\
    A.~Faraggi and L.~A.~P.~Zayas,
    ``The Spectrum of Excitations of Holographic Wilson Loops,''
    arXiv:1101.5145 [hep-th], \\
 E.~I.~Buchbinder and A.~A.~Tseytlin,
  ``The 1/N correction in the D3-brane description of circular Wilson loop at strong coupling,''
  arXiv:1404.4952 [hep-th].


\bibitem{WLref}
N.~Drukker, D.~J.~Gross and H.~Ooguri,
  ``Wilson loops and minimal surfaces,''
  Phys.\ Rev.\  D {\bf 60}, 125006 (1999)
  [arXiv:hep-th/9904191], \\
  N.~Drukker, S.~Giombi, R.~Ricci and D.~Trancanelli,
  ``Supersymmetric Wilson loops on S**3,''
  JHEP {\bf 0805}, 017 (2008)
  [arXiv:0711.3226 [hep-th]], \\
  N.~Drukker and B.~Fiol,
  ``On the integrability of Wilson loops in AdS(5) x S**5: Some periodic
  ansatze,''
  JHEP {\bf 0601}, 056 (2006)
  [arXiv:hep-th/0506058], \\
    K.~Zarembo,
    ``Supersymmetric Wilson loops,''
    Nucl.\ Phys.\  B {\bf 643}, 157 (2002)
    [arXiv:hep-th/0205160],
     N.~Drukker, S.~Giombi, R.~Ricci and D.~Trancanelli,
                         ``Supersymmetric Wilson loops on S**3,''
                         JHEP {\bf 0805}, 017 (2008)
                         [arXiv:0711.3226 [hep-th]].
 

\bibitem{cusp}
M.~Kruczenski,
  ``A note on twist two operators in N = 4 SYM and Wilson loops in Minkowski
signature,''
  JHEP {\bf 0212}, 024 (2002)
  [arXiv:hep-th/0210115].

\bibitem{scatampl}
See e.g. \\
 L.~F.~Alday and J.~M.~Maldacena,
  ``Gluon scattering amplitudes at strong coupling,''
  JHEP {\bf 0706}, 064 (2007)
  [arXiv:0705.0303 [hep-th]], \\ 
    L.~F.~Alday and J.~Maldacena,
    ``Null polygonal Wilson loops and minimal surfaces in Anti-de-Sitter space,''
    JHEP {\bf 0911}, 082 (2009)
    [arXiv:0904.0663 [hep-th]], \\
  L.~F.~Alday and J.~Maldacena,
  ``Comments on gluon scattering amplitudes via AdS/CFT,''
  JHEP {\bf 0711}, 068 (2007)
  [arXiv:0710.1060 [hep-th]], \\
  J.~Maldacena and A.~Zhiboedov,
  ``Form factors at strong coupling via a Y-system,''
  JHEP {\bf 1011}, 104 (2010)
  [arXiv:1009.1139 [hep-th]], \\
  L.~F.~Alday, B.~Eden, G.~P.~Korchemsky, J.~Maldacena and E.~Sokatchev,
  ``From correlation functions to Wilson loops,''
  arXiv:1007.3243 [hep-th], \\
  L.~F.~Alday, D.~Gaiotto, J.~Maldacena, A.~Sever and P.~Vieira,
  ``An Operator Product Expansion for Polygonal null Wilson Loops,''
  arXiv:1006.2788 [hep-th],  \\
  L.~F.~Alday, J.~Maldacena, A.~Sever and P.~Vieira,
  ``Y-system for Scattering Amplitudes,''
  J.\ Phys.\ A  {\bf 43}, 485401 (2010)
  [arXiv:1002.2459 [hep-th]], \\
  H.~Dorn, N.~Drukker, G.~Jorjadze and C.~Kalousios,
        ``Space-like minimal surfaces in AdS x S,''
        JHEP {\bf 1004}, 004 (2010)
        [arXiv:0912.3829 [hep-th]].





\bibitem{IKZ} 
  R.~Ishizeki, M.~Kruczenski and S.~Ziama,
  ``Notes on Euclidean Wilson loops and Riemann Theta functions,''
  Phys.\ Rev.\ D {\bf 85}, 106004 (2012)
  [arXiv:1104.3567 [hep-th]].

  
\bibitem{KZ} 
  M.~Kruczenski and S.~Ziama,
  ``Wilson loops and Riemann theta functions II,''
  JHEP {\bf 1405}, 037 (2014)
  [arXiv:1311.4950 [hep-th]].

\bibitem{BB}
M.~Babich and A.~Bobenko,
``Willmore Tori with umbilic lines and minimal surfaces in hyperbolic space'',
 Duke Mathematical Journal {\bf 72, No. 1}, 151 (1993).

\bibitem{BBook}
E.~D.~Belokolos, A.~I.~Bobenko,V.~Z.~Enol'skii, A.~R.~Its, V.~B.~Matveev,
``Algebro-Geometric Approach to Nonlinear Integrable Equations,''
Springer-Verlag series in Non-linear Dynamics,
Springer-Verlag Berlin Heidelberg NewYork (1994).

\bibitem{ClosedStrings}
  A.~Jevicki and K.~Jin,
  ``Moduli Dynamics of AdS(3) Strings,''
  JHEP {\bf 0906}, 064 (2009)
  [arXiv:0903.3389 [hep-th]],\\
  A.~Jevicki, K.~Jin, C.~Kalousios and A.~Volovich,
           ``Generating AdS String Solutions,''  
           JHEP {\bf 0803}, 032 (2008)
           [arXiv:0712.1193 [hep-th]], \\   
            M.~Kruczenski,
             ``Spin chains and string theory,''
             Phys.\ Rev.\ Lett.\  {\bf 93}, 161602 (2004)
             [arXiv:hep-th/0311203], \\       
M.~Kruczenski,
  ``Spiky strings and single trace operators in gauge theories,''
  JHEP {\bf 0508}, 014 (2005)
  [arXiv:hep-th/0410226], \\
 N.~Dorey and B.~Vicedo,
  ``On the dynamics of finite-gap solutions in classical string theory,''
  JHEP {\bf 0607}, 014 (2006)
  [arXiv:hep-th/0601194], \\
  K.~Sakai and Y.~Satoh,
  ``Constant mean curvature surfaces in \ads{3},''
  JHEP {\bf 1003}, 077 (2010)
  [arXiv:1001.1553 [hep-th]], \\
H.~J.~De Vega and N.~G.~Sanchez,
          ``Exact integrability of strings in D-Dimensional De Sitter space-time,''
          Phys.\ Rev.\ D {\bf 47}, 3394 (1993), \\
             A.~L.~Larsen and N.~G.~Sanchez,
             ``Sinh-Gordon, cosh-Gordon and Liouville equations for strings and multistrings in constant curvature space-times,''  
             Phys.\ Rev.\ D {\bf 54}, 2801 (1996)  [hep-th/9603049], \\
K.~Zarembo,
                               ``Wilson loop correlator in the AdS / CFT correspondence,''
                               Phys.\ Lett.\ B {\bf 459}, 527 (1999)
                               [hep-th/9904149], \\
 N.~Drukker and B.~Fiol,
           ``On the integrability of Wilson loops in AdS(5) x S**5: Some periodic
           ansatze,''
           JHEP {\bf 0601}, 056 (2006)
           [arXiv:hep-th/0506058].


\bibitem{Janik} 
  R.~A.~Janik and P.~Laskos-Grabowski,
  ``Surprises in the AdS algebraic curve constructions: Wilson loops and correlation functions,''
  Nucl.\ Phys.\ B {\bf 861}, 361 (2012)
  [arXiv:1203.4246 [hep-th]].

\bibitem{WLint}
               B.~Fiol and G.~Torrents,
               ``Exact results for Wilson loops in arbitrary representations,''
               arXiv:1311.2058 [hep-th], \\
                  D.~Muller, H.~Munkler, J.~Plefka, J.~Pollok and K.~Zarembo,
                  ``Yangian Symmetry of smooth Wilson Loops in N=4 super Yang-Mills Theory,''
                  arXiv:1309.1676 [hep-th], \\
                  S.~Ryang,
                  ``Algebraic Curves for Long Folded and Circular Winding Strings in AdS5xS5,''
                  JHEP {\bf 1302}, 107 (2013)
                  [arXiv:1212.6109 [hep-th]], \\
                                  A.~Dekel,
                                  ``Algebraic Curves for Factorized String Solutions,''
                                  JHEP {\bf 1304}, 119 (2013)
                                  [arXiv:1302.0555 [hep-th]], \\
                                  A.~Dekel and T.~Klose,
                                    ``Correlation Function of Circular Wilson Loops at Strong Coupling,''
                                    JHEP {\bf 1311}, 117 (2013)
                                    [arXiv:1309.3203 [hep-th]], \\
                  A.~Irrgang and M.~Kruczenski,
                  ``Rotating Wilson loops and open strings in AdS3,''
                  J.\ Phys.\ A {\bf 46}, 075401 (2013)
                  [arXiv:1210.2298 [hep-th]], \\
                  A.~Irrgang and M.~Kruczenski,
                                                       ``Double-helix Wilson loops: Case of two angular momenta,''
                                                       JHEP {\bf 0912}, 014 (2009)
                                                       [arXiv:0908.3020 [hep-th]], \\
                  V.~Forini, V.~G.~M.~Puletti and O.~Ohlsson Sax,
                  ``Generalized cusp in $AdS_4 \times CP^3$ and more one-loop results from semiclassical strings,''
                  J.\ Phys.\ A {\bf 46}, 115402 (2013)
                  [arXiv:1204.3302 [hep-th]], \\
                    B.~A.~Burrington and L.~A.~Pando Zayas,
                    ``Phase transitions in Wilson loop correlator from integrability in global AdS,''
                    Int.\ J.\ Mod.\ Phys.\ A {\bf 27}, 1250001 (2012)
                    [arXiv:1012.1525 [hep-th]], \\
               G.~Papathanasiou,
               ``Pohlmeyer reduction and Darboux transformations in Euclidean worldsheet $AdS_3$,''
               JHEP {\bf 1208}, 105 (2012)
               [arXiv:1203.3460 [hep-th]], \\
               N.~Drukker and V.~Forini,
               ``Generalized quark-antiquark potential at weak and strong coupling,''
               JHEP {\bf 1106}, 131 (2011)
               [arXiv:1105.5144 [hep-th]], \\
               B.~A.~Burrington,
               ``General Leznov-Savelev solutions for Pohlmeyer reduced AdS$_5$ minimal surfaces,''
               JHEP {\bf 1109}, 002 (2011)
               [arXiv:1105.3227 [hep-th]], \\
               L.~F.~Alday and A.~A.~Tseytlin,
               ``On strong-coupling correlation functions of circular Wilson loops and local operators,''
               J.\ Phys.\ A {\bf 44}, 395401 (2011)
               [arXiv:1105.1537 [hep-th]], \\
               C.~Kalousios and D.~Young,
               ``Dressed Wilson Loops on $S^2$,''
               Phys.\ Lett.\ B {\bf 702}, 299 (2011)
               [arXiv:1104.3746 [hep-th]],\\
                R.~Ishizeki, M.~Kruczenski and A.~Tirziu,
                 ``New open string solutions in AdS(5),''
                 Phys.\ Rev.\ D {\bf 77}, 126018 (2008)
                 [arXiv:0804.3438 [hep-th]].
           
\bibitem{Cagnazzo} 
  A.~Cagnazzo,
  ``Integrability and Wilson loops: the wavy line contour,''
  arXiv:1312.6891 [hep-th].
  
             
 \bibitem{Pohlmeyer}
 K. Pohlmeyer, 
 ``Integral Hamiltonian systems and interactions through quadratic constraints,'' 
Commun. Math. Phys. 46, 207 (1976). 
           
\bibitem{Hoare:2012nx} 
                   B.~Hoare and A.~A.~Tseytlin,
                   ``Pohlmeyer reduction for superstrings in AdS space,''
                   arXiv:1209.2892 [hep-th].

\bibitem{SARM}
 S.~Alexakis and R.~Mazzeo,
  ``Renormalized area and properly embedded minimal surfaces in hyperbolic
  3-manifolds,''
  Commun.\ Math.\ Phys.\  {\bf 297}, 621 (2010).

\bibitem{Hill}
Wilhelm Magnus, Stanley Winkler , 
 "Hill's Equation", Dover Books on Mathematics, Dover Publications (2004).
           
\bibitem{SY} 
  G.~W.~Semenoff and D.~Young,
  ``Wavy Wilson line and AdS / CFT,''
  Int.\ J.\ Mod.\ Phys.\ A {\bf 20}, 2833 (2005)
  [hep-th/0405288].

\bibitem{CHMS} 
  D.~Correa, J.~Henn, J.~Maldacena and A.~Sever,
  ``An exact formula for the radiation of a moving quark in N=4 super Yang Mills,''
  JHEP {\bf 1206}, 048 (2012)
  [arXiv:1202.4455 [hep-th]].
 
 
 \bibitem{Drukker} 
   N.~Drukker,
   ``1/4 BPS circular loops, unstable world-sheet instantons and the matrix model,''
   JHEP {\bf 0609}, 004 (2006)
   [hep-th/0605151].
   
\bibitem{Marchenko}
 V. A. Marchenko, Sturm-Liouville operators and their applications “Naukova Dumka", Kiev, 1977; English transl., Birkhauser, 1986.   
   
\bibitem{ThF}
D.~Mumford, (with the collaboration of C.~Musili, M.~Nort,E.~Previato and M.~Stillman),
``Tata Lectures in Theta I \& II'', 
Modern Birkh\"auser Classics, Birkh\"auser, Boston (2007), \\
John D. Fay, 
 ``Theta Functions on Riemann Surfaces'', 
Lectures Notes in Mathematics {\bf 352},Springer-Verlag, Berlin Heidelberg, New York (1973),\\
H.~F.~Baker, 
``Abel's Theorem and the Allied Theory, Including the Theory of the Theta Functions'',
Cambridge University Press (1897).
 
\bibitem{FK}
 H.~M.~Farkas and I.~Kra,
 ``Riemann Surfaces'',
 Graduate Texts in Mathematics, Second Edition, Springer-Verlag, New, Berlin, Heidelberg (1991).

\bibitem{JD}
Jesse Douglas, ``Solution of the problem of Plateau'', Transactions of the American Mathematical Society, Vol. 33, (1931), 263-321.


\bibitem{Ambjorn} 
  J.~Ambjorn and Y.~Makeenko,
  Phys.\ Rev.\ D {\bf 85}, 061901 (2012)
  [arXiv:1112.5606 [hep-th]].

\bibitem{Toledo} J. Toledo, in preparation.
           
           
           
           
           
           
 





\end{thebibliography}
\end{document}